\documentclass[reprint,prl]{revtex4-1}
\usepackage{amssymb,amsmath}
\usepackage{graphicx}
\usepackage{dcolumn}
\usepackage{multirow}
\usepackage{color}
\usepackage{bm}

\newcommand{\ket}[1]{|{#1}\rangle}

\def\simlt{\mathrel{\lower .3ex \rlap{$\sim$}\raise .5ex \hbox{$<$}}}
\def\simgt{\mathrel{\lower .3ex \rlap{$\sim$}\raise .5ex \hbox{$>$}}}

\newcommand{\snc}[1]{\textcolor{black}{#1}}
\newcommand{\mae}[1]{\textcolor{black}{#1}}
\newcommand{\lz}[1]{\textcolor{black}{#1}}

\begin{document}

\title{\textbf{\fontfamily{phv}\selectfont Two-axis control of a singlet-triplet qubit with an integrated micromagnet}}
\author{Xian Wu, D. R. Ward, J. R. Prance, Dohun Kim, John King Gamble, R.T. Mohr, Zhan Shi, D. E. Savage, M. G. Lagally, Mark Friesen, S. N. Coppersmith, and M. A. Eriksson
\\University of Wisconsin-Madison, Madison, WI 53706}

\begin{abstract}
The qubit is the fundamental building block of a quantum computer.
We fabricate a qubit in a silicon double quantum dot with
an integrated micromagnet in which the qubit basis states are the
singlet state and the spin-zero triplet state of two electrons.
Because of the micromagnet,
the magnetic field difference $\bm{\Delta B}$ between the two sides
of the double dot is large enough to enable
the achievement of coherent rotation of the qubit's Bloch vector
about two different axes of the Bloch sphere.
By measuring the decay of the quantum oscillations, the
inhomogeneous spin coherence time $\bm{T_{2}^{*}}$ is determined.
By measuring $\bm{T_{2}^{*}}$ at many different values of the exchange coupling $\bm{J}$ and at two different values of $\bm{\Delta B}$, we provide evidence that the micromagnet does not limit decoherence, with
the dominant limits on $\bm{T_{2}^{*}}$ arising from charge noise and from coupling to nuclear spins. 
 
\end{abstract}

\maketitle

Fabricating qubits composed of 
electrons in semiconductor quantum dots is a promising approach for the
development of a large-scale quantum computer because of
the approach's potential for scalability and for integrability with classical electronics.  
Much recent progress
has been made, and spin manipulation has been demonstrated in systems of 
two~\cite{Levy:2002p1446,Petta:2005p2180,Nowack:2011p1269,Shi:2014p3020,Kim:2014p1401.4416},
three~\cite{Gaudreau:2011p54,Medford:2013p654}, and four~\cite{Shulman:2012p202} quantum dots.  A great deal of attention has focused on the singlet-triplet qubit in quantum dots~\cite{Levy:2002p1446, Petta:2005p2180, Reilly:2008p817, Foletti:2009p903, Barthel:2009p160503, Barthel:2010p266808, Bluhm:2011p109, Maune:2012p344, Shi:2011p233108, Otsuka:2012p081308, Studenikin:2012p226802, Dial:2013p146804}, which consists of 
the $S_z=0$ subspace of two electrons, for which the
basis can be chosen to be a singlet and a triplet state.
Full two-axis control on the Bloch sphere is achieved by
 electrical gating in the presence of a magnetic field difference $\Delta B$ between the two dots.
 In previous experiments~\cite{Petta:2005p2180,Reilly:2008p817,Foletti:2009p903,Barthel:2009p160503,Barthel:2010p266808,Bluhm:2011p109,Maune:2012p344}, $\Delta B$ arises from coupling to nuclear spins in the 
 material, and slow fluctuations in these nuclear fields lead to inhomogeneous decoherence times
 that, without special nuclear state preparation, typically are  shorter than the period of the quantum oscillations.
 In III-V materials, $\Delta B$ is large, so fast oscillation periods of order 10~ns are achievable, 
 but the inhomogeneous dephasing time is
also $\sim$10~ns, so that oscillations from $\Delta B$ are overdamped, ending before a complete cycle is observed~\cite{Petta:2005p2180}.
 The fluctuations of the nuclear spin bath can be mitigated to some extent~\cite{Foletti:2009p903},
 but inhomogeneous dephasing times in III-V materials are short enough that high-fidelity control is still very challenging.
 Coupling to nuclear spins in silicon is substantially weaker, leading to longer coherence times,
 but also smaller field differences and hence slower quantum oscillations~\cite{Maune:2012p344,Zwanenburg:2013p961}.

Here, we report the operation of a singlet-triplet qubit in which the magnetic field difference $\Delta B$ between
the dots is imposed by an external micromagnet~\cite{PioroLadriere:2007p024105, PioroLadriere:2008p776}. 
Because the field from the micromagnet is stable in time, a large $\Delta B$ can be imposed without
creating inhomogeneous dephasing.  
We present data demonstrating underdamped quantum oscillations, and, by investigating a variety of voltage configurations and two $\Delta B$ configurations, we show that the micromagnet indeed increases $\Delta B$ without significantly increasing inhomogeneous dephasing rates induced by coupling to nuclear spins.

\begin{figure*}
\begin{center}
\includegraphics[width=\textwidth]{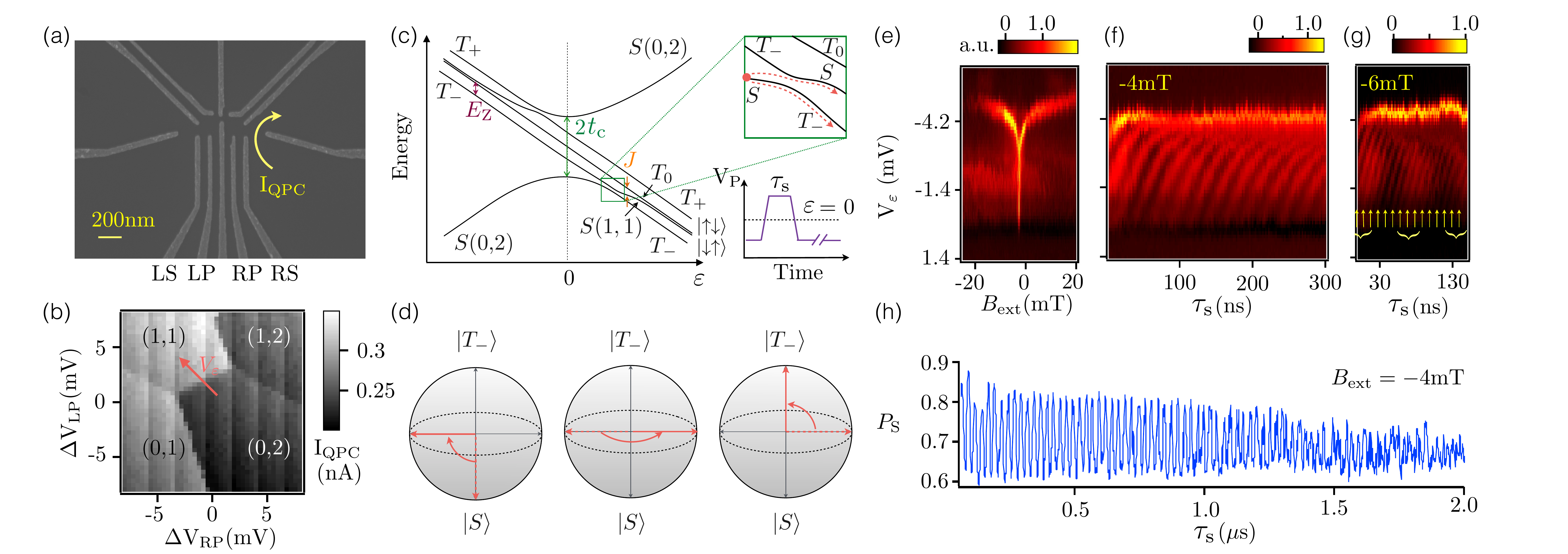}
\caption{\label{fig1}(a) Scanning electron micrograph of a device identical to the one used in the
experiment before deposition of the gate dielectric and accumulation gates. {\color{black} An optical image of a complete device showing the micromagnet is included in the SI Appendix.}
Gates labelled LS and RS are used for fast pulsing. The curved arrow shows the current path through the quantum point contact (QPC) used as a charge sensor.  
(b) $I_\mathrm{QPC}$ measured as a function of $V_\mathrm{LP}$ and $V_\mathrm{RP}$ yields the double-dot charge stability diagram. Electron numbers in the left and right dot are indicated on the diagram.
The red arrow denotes the direction in gate voltage space {\color{black}$V_{\varepsilon} =\sqrt{\Delta V_{LP}^{2}+\Delta V_{RP}^{2}}$} that changes the detuning $\varepsilon$ between the quantum dots.
(c) Schematic energy diagram near the (0,2)-to-(1,1) charge transition, showing energies of singlet $S$ and triplet $T$ states as functions of $\varepsilon$. The exchange energy splitting $J$ between $S$ and $T_{0}$, the Zeeman splitting $E_\mathrm{Z}$ between $T_{-}$ and $T_{0}$, and the tunnel coupling $t_\mathrm{c}$ are also shown. At large $\varepsilon$, in the presence of a field difference between the
two dots, $S$ and $T_0$ mix, and the corresponding energy eigenstates are $\ket{\uparrow\downarrow}$ and $\left |\downarrow\uparrow\right\rangle$. At small $\varepsilon$, the small transverse field from the micromagnet and the nuclear fields turns the $S$-$T_{-}$ crossing into an anti-crossing (zoom in). Pulsing through this anti-crossing with intermediate velocity transforms $S$ into a superposition of $S$ and  $T_-$,
leading to Landau-St$\mathrm{\ddot{u}}$ckelberg-Zener oscillations at the frequency corresponding to the $S$-$T_{-}$
energy difference~\cite{Petta:2010p669}.  
The pulse used to observe the spin funnel and $S$-$T_{-}$ oscillations \color{black}{shown in (e)} is also shown, where the pulse voltage $V_\mathrm{P}$ is applied along the detuning axis.
(d) Bloch sphere representation of $\pi$ rotation of $S$ and $T_{-}$ states with $50\%$ initialization into each state.
(e) Spin funnel~\cite{Petta:2005p2180} measurement of the location of the $S$-$T_{-}$ anti-crossing as a function of external magnetic field $B_\mathrm{ext}$ and $V_{\varepsilon}$. The data were acquired by sweeping along the detuning direction with the pulse on, with the vertical axis reporting the value of the detuning
at the base of the pulse. 
The spin funnel occurs when $S$-$T_{-}$ mixing is fast, which locates the relevant anticrossing.
(f),(g) $S$-$T_{-}$ oscillations acquired at different external $B$ fields. The oscillation frequency increases with increasing $B_\mathrm{ext}$. {\color{black}The slower oscillations in (g) with period $\sim 80$~ns and labelled with the curly brackets are $S$-$T_{0}$ oscillations, which are investigated in more detail in Fig.\ 3. \mae{The $S$-$T_{-}$ oscillations in (g) are labelled with arrows.}}
(h) Singlet probability as a function of pulse duration $\tau_\mathrm{s}$ at external magnetic field $B$ = -4~mT and base detuning {\color{black}$V_{\varepsilon} \simeq -2.8$~mV.}}
\end{center}
\end{figure*}

A top view of the double quantum dot device\snc{, which is fabricated in a Si/SiGe heterostructure,} is shown in Fig.\ 1(a);
\snc{fabrication techniques}
are discussed in the Materials and Methods, \mae{and an optical image of the micromagnet can be found in the SI Appendix}.
The charge occupation of the two sides of the double dot is determined by measuring
the current through a quantum point contact (QPC) next to one of the dots, as shown in Fig.\ 1(a).
Fig.\ \ref{fig1}(b) shows a charge stability diagram, obtained by measuring the current through the quantum point contact (QPC) as a function of gate voltages on LP and RP; the number of electrons on each side of the dot is labelled.
The qubit manipulations are performed in the (1,1) region (detuning $\varepsilon > 0$), while initialization and readout are carried out in the (0,2) region ($\varepsilon < 0$). 
Fig.\ \ref{fig1}(c) shows the energy level diagram at small but nonzero magnetic field. The three triplet states $T_- = |{\downarrow\downarrow}\rangle $,
$T_0 = (|{\uparrow\downarrow}\rangle+|{\downarrow\uparrow}\rangle)/\sqrt{2}$, and
$T_+=|{\uparrow\uparrow}\rangle$ are split from each other  by the Zeeman energy $E_\mathrm{Z} = g \mu_\mathrm{B} B_\mathrm{ave}$, where $g$ is the gyromagnetic ratio, $\mu_\mathrm{B}$ is the Bohr magneton, and $B_\mathrm{ave}$ is the average of the total magnetic field. A difference in the transverse magnetic fields on the dots, either from the external micromagnet or
from nuclear hyperfine fields, mixes the singlet $S$ and triplet $T_{-} $ states and turns the $S$-$T_{-}$ crossing into an anti-crossing.
This avoided crossing enables the observation of a spin funnel {\color{black}{where
the $S$-$T_{-}$ mixing is fast}}~\cite{Petta:2005p2180} as well as quantum oscillations between $S$ and $T_{-}$~\cite{Petta:2010p669}. The spin funnel is shown in Fig.\ \ref{fig1}(e), and the $S$-$T_{-}$ oscillations are shown in Fig.\ \ref{fig1}(f,g). The applied pulse in Fig.\ \ref{fig1}(e) is a simple one-stage pulse along the detuning direction with fixed amplitude, repeated at a rate of 33\ kHz, which is slow enough for spin relaxation to reinitialize to the singlet before application of the next pulse~\cite{Prance:2012p046808,Shi:2012p140503}. 
{\color{black}The lever arm $\alpha$, the conversion between detuning energy $\varepsilon$ and gate voltage $V_{\varepsilon}$, is 35.4$\ \mu$eV/mV. See the SI Appendix for methods used to extract $\alpha$ and convert the measured QPC current to the probability of being in the $S$ state at the end of the applied pulse.}
The spin funnel is obtained by sweeping along the detuning direction (i.e., sweeping {\color{black}$V_\varepsilon$}) with the pulse on,
and stepping the external magnetic field $B_\mathrm{ext}$. 
When the pulse tip reaches the $S$-$T_{-}$ anti-crossing, a strong resonance signal is observed, corresponding to strong mixing of $S$-$T_{-}$ states. Since right at the anti-crossing $E_\mathrm{Z} \simeq J$, we can map out $J$ at small $\varepsilon$ by sweeping the magnetic field. 
{\color{black}{The center of the spin funnel occurs when the applied field cancels out the average field from the micromagnet, which indicates $B_\mathrm{ave}\simeq2.5$~mT.}}
The tunnel coupling $t_{c}\sim 3.4\ \mu$eV is estimated from the dependence of the location of the spin funnel on magnetic field~\cite{Petta:2005p2180}. {\color{black} The pulse rise time of 10~ns ensures nearly adiabatic passage over the S(0,2)-S(1,1) anti-crossing, with 
a non-adiabatic transition probability $<0.1\%$\cite{Shevchenko:2010p1}. }

{\color{black} By 
increasing the rise time
of the pulse, so that it is slower than that used to observe the spin funnel, the voltage pulse can be used to cause $S$ to evolve into a superposition of the $S$ and $T_{-}$ states.  In this case, the pulse remains adiabatic with respect to the S(0,2)-S(1,1) anti-crossing; it is, however, only quasi-adiabatic with respect to the $S$-$T_{-}$ anticrossing, enabling use of the Landau-Zener mechanism to initialize a superposition between states $S$ and $T_{-}$ (see Fig.\ \ref{fig1}(c), inset)~\cite{Petta:2010p669, Ribeiro:2013p235318, Cao:2013p1401,Granger2014preprint}.  As the voltage pulse takes these states to larger detuning, an energy difference arises between the pair of states, and there is a relative phase accumulation between them.  The return pulse leads to quantum interference between these two states and to oscillations in the charge occupation  as a function of the acquired phase.}
Fig.~\ref{fig1}(d) illustrates the ideal case, in which the rising edge of the pulse transforms $S$ 
into an equal superposition of  $S$ and $T_{-}$, followed by accumulation of a relative phase difference of $\pi$ after pulse duration $\tau_{S}$. Fig.\ \ref{fig1}(f) shows $S$-$T_{-}$ oscillations at $B_\mathrm{ext}=-4$~mT, obtained by applying a pulse with a rise time of 45 ns.
\mae{Fig.\ 1(h) reports a line scan of the singlet probability for $S$-$T_{-}$ oscillations measured at $B_\mathrm{ext}=-4$~mT;} for this measurement the tip of the voltage pulse reaches large enough detuning that $E_{ST_{-}}$ is essentially constant and independent of detuning. 
\mae{From this data we extract a dephasing time of $1.7~\mathrm{\mu s}$ by fitting the oscillation amplitude to a gaussian decay function of the pulse duration $\tau_{S}$. The $S$-$T_{-}$ oscillations observed here are longer-lived  than those observed in GaAs~\cite{Petta:2010p669}, presumably in part because Si has weaker hyperfine fields~\cite{Assali:2011p165301}.}  \lz{However, the visibility here is similar to that in GaAs, indicating that decoherence is still important in limiting the ability to tune the pulse rise time to achieve equal amplitude in the $S$ and $T_{-}$ branches of the Landau-Zener beam splitter~\cite{Petta:2010p669, Ribeiro:2013p235318, Cao:2013p1401,Granger2014preprint}.}
\mae{Fig.\ \ref{fig1}(g) shows a similar measurement for which we used a slightly faster (16~ns) rise time for the pulse, the effect of which is to increase the overlap of the wavefunction with the singlet state $S$.  As a result, both $S$-$T_{-}$ oscillations and $S$-$T_{0}$ oscillations are visible in this plot, which was acquired at $B_\mathrm{ext}=-6$~mT. 
The faster oscillations with period 10~ns, marked with the small arrows in Fig.\ 1(g), are the $S$-$T_{-}$ oscillations. 
The slower oscillations, marked with the curly brackets, are the $S$-$T_{0}$ oscillations.  As we discuss below, these latter oscillations can be made dominant by further modifications of the manipulation pulse, and for these oscillations the micromagnet plays a critical role in enhancing the rotation rate on the $S$-$T_{0}$ Bloch sphere.}

\begin{figure*}[t]
\begin{center}
\includegraphics[width=\textwidth]{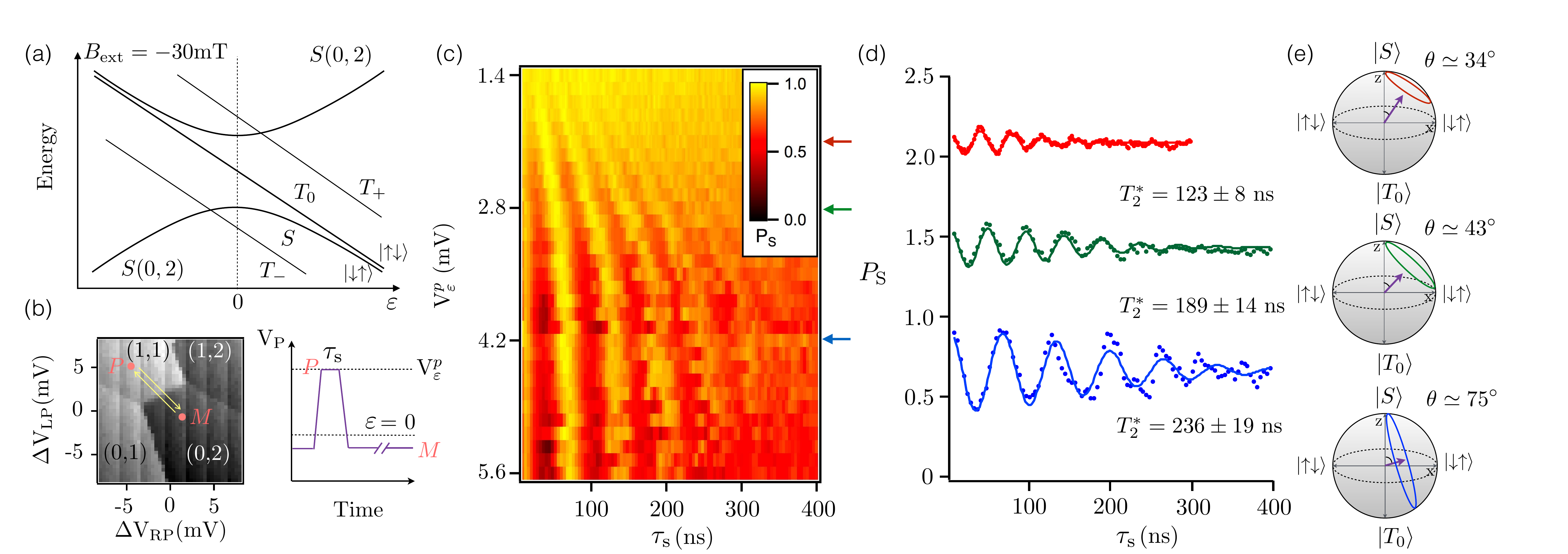}
\caption{\label{fig2}(a) Schematic energy level diagram near the (0,2)-(1,1) charge transition at
external field
 $B_\mathrm{ext}=-30$mT. 
(b) Pulse sequence used to observe $S$-$T_{0}$ oscillations. Starting at point M in the $S$(0,2) ground state, a fast adiabatic pulse into (1,1) is applied (it is adiabatic for the $S$(0,2)-$S$(1,1) anticrossing and
sudden for the $S$(1,1)-$T_0$ anticrossing), to point P, where the exchange coupling $J$ 
is comparable to or less than $h$, the energy from the magnetic field difference. The speed and axis of
the rotation on the Bloch sphere during the pulse of duration $\tau_\mathrm{S}$ depend on both $J$ and $h$. 
 Readout is performed by reversing the fast adiabatic pulse, which converts
$S$(1,1) to $S$(0,2) but does not change the charge configuration of $T_0$.
(c) Probability $P_{S}$ of being in state $S$ as a function of detuning $\varepsilon$ and pulse duration 
$\tau_\mathrm{s}$. Here, the measurement point M in the (0,2) charge state is fixed and the detuning at
the pulse tip, $V_\mathrm{\varepsilon}^\mathrm{p}$, and pulse duration, $\tau_\mathrm{s}$, are varied. 
(d) $P_\mathrm{S}$ as a function of $\tau_\mathrm{s}$, extracted from the data in (c) at three different values of $V_\mathrm{\varepsilon}$. Each trace is fit to the product of a cosine and a gaussian~\cite{Petersson:2010p246804,Beaudoin:2013p085320}, with amplitude, frequency, phase and decay time as free parameters (solid curves). The decay time $T_{2}^{*}$ is listed for each trace. Each trace is offset by 0.6 for clarity.
(e) Bloch spheres showing the rotations corresponding to each trace in (d).  The angle $\theta$  between rotation axis and the $z$-axis is labeled for each case. } 
\end{center}
\end{figure*}

We investigate the S-T$_{0}$ oscillations, which correspond to a gate rotation of the S-T$_{0}$
qubit, in more detail by changing the applied magnetic field $B_\mathrm{ext}$ to -30~mT, \mae{and by working with faster pulse rise times.}  Here the $S$-$T_{-}$ anticrossing occurs at negative $\varepsilon$, as shown in Fig.\ 2(a), making it easier to 
pulse through that anticrossing
quickly enough so that the state remains $S$.  In this situation, the relevant Hamiltonian $H$ for $\epsilon > 0$, in the $S$-$T_{0}$ basis, is

\begin{equation}
H=\left[\begin{array}{cc}-J(\epsilon) & h/2 \\h/2 & 0\end{array}\right].\\[6pt]
\end{equation}
\noindent
Here, $J$ is the exchange coupling, and $h=g\mu_\mathrm{B}\Delta B$ is the energy contribution from the magnetic field difference. The angle $\theta$ between the rotation axis and the $z$-axis of the Bloch sphere satisfies $\tan{\theta} = h/J$, and the rotation angular frequency $\omega = \sqrt{h^{2}+J^{2}}/\hbar$.
Both $\theta$ and $\omega$ depend on $\varepsilon$, because $J$ varies with $\varepsilon$.

Rotations about the $x$-axis of the Bloch sphere (the ``$\Delta B$ gate'') are implemented using
the simple one stage pulse shown in Fig.\ \ref{fig2}(b), starting from point M in the (0,2) charge state.
The pulse rise time of a few ns is slow enough that the pulse is adiabatic through the $S$(0,2)-$S$(1,1) anticrossing.  \mae{As $\varepsilon$ increases, the eigenstates transition from $S$(1,1) and $T_0$ to other combinations of $\left |\uparrow\downarrow\right\rangle$ and $\left |\downarrow\uparrow\right\rangle$, and in the limit of $\varepsilon \rightarrow \infty$, the eigenstates become $\left |\uparrow\downarrow\right\rangle$ and $\left |\downarrow\uparrow\right\rangle$.}  \mae{The voltage pulse applied is sudden with respect to this transition in the energy eigenstates, so that, immediately following the rising edge of the pulse, the system remains in $S$(1,1).}
\mae{At large detuning, $J \le h$,  and $S$-$T_{0}$ oscillations are observed following the returning edge of the pulse.}  \mae{These oscillations arise from the $x$-component of the rotation axis and have a rotation rate that is largely determined by the magnitude of $h$.} Fig.\ \ref{fig2}(c) shows the singlet probability $P_{S}$ plotted as a function of the detuning voltage at the pulse tip, $V_\mathrm{\varepsilon}^\mathrm{p}$, and pulse duration $\tau_{S}$.  
\mae{The data in the top 1/3 of the figure were acquired with a pulse rise time of 2.5~ns, and the data shown in the bottom 2/3 of the figure were acquired using a 5~ns rise time.}
As is clear from Fig.~2(c,d), $J$ decreases as $\varepsilon$ increases, so the oscillation angular frequency becomes smaller and approaches $h/\hbar$ as $J \rightarrow 0$. 
The visibility of the oscillations is largest at large $V_\mathrm{\varepsilon}^\mathrm{p}$, because in that regime the rotation axis is closest to the $x$-axis, as shown in Fig.\ \ref{fig2}(e). By fitting traces from Fig.\ \ref{fig2}(c)  to the product of a cosine and a gaussian~\cite{Petersson:2010p246804}, we extract the inhomogeneous dephasing time $T_{2}^{*}$ as a function of $\varepsilon$.  
\mae{Based on the rotation period at large $\varepsilon$, we estimate $h \approx 60.5$~neV, which corresponds to an X-rotation rate of 14\ MHz.  The rotation rate we observe here is much faster than the X-rotation rate achievable without micromagnets in Si, which is 460 kHz~\cite{Maune:2012p344}; micromagnets closer to the quantum dots offer the potential for even faster rotation rates than those reported here.  Using feedback to prepare the nuclear spins in GaAs quantum dots, X-rotation rates of 30 MHz rates have been reported~\cite{Dial:2013p146804}, comparable but slightly faster than the rates we achieve here without such preparation.} 
 
\begin{figure*}[t]
\begin{center}
\includegraphics[width=\textwidth]{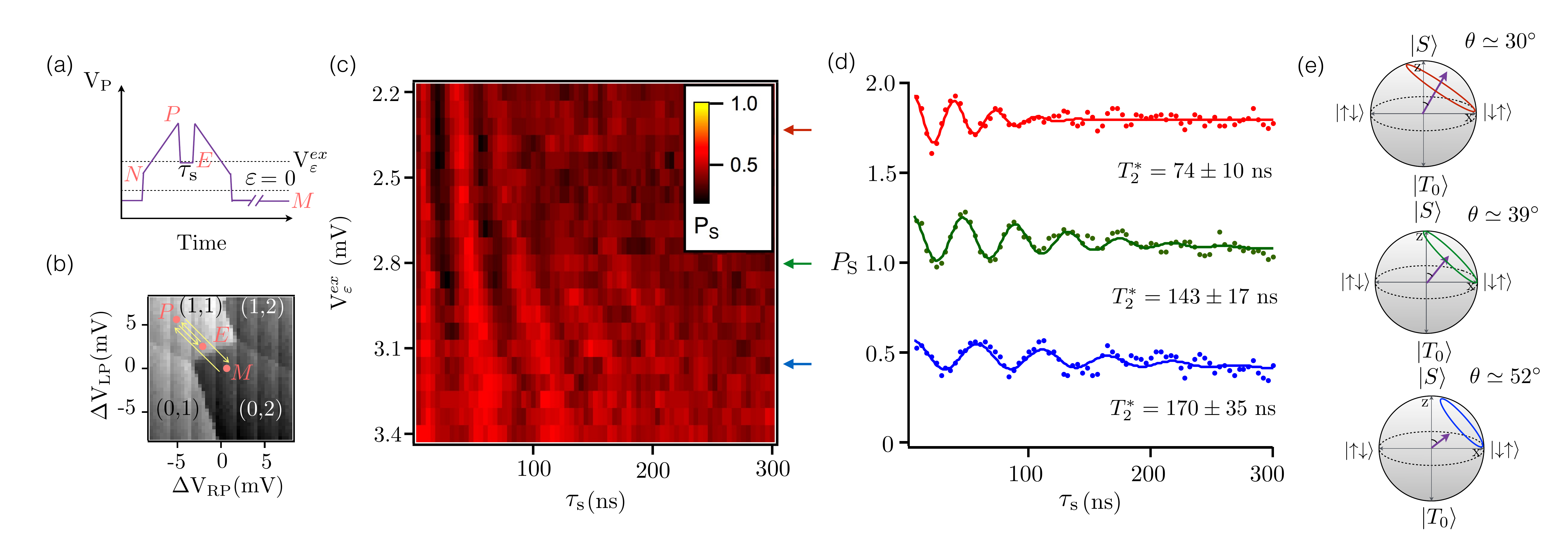}
\caption{\label{fig3}(a-b) Pulse sequence used to observe $S$-$T_{0}$ oscillations when $J > h$. We initialize into the $S$(1,1) state by preparing the $S$(0,2) ground state at point M and ramping adiabatically through the (0,2)-(1,1)$S$ anti-crossing to an intermediate point N and then to P, where the singlet and triplet states are no longer energy eigenstates.  Decreasing $\varepsilon$ suddenly brings the state non-adiabatically to a value of the detuning where $J$ is comparable or greater than $h$, inducing coherent rotations. The Bloch vector rotates around the new axis for a time $\tau_\mathrm{s}$. Reversing the sequence of ramps projects the state into $S$(0,2) for readout. 
(c) Probability $P_{S}$ of observing the singlet as a function of the detuning of the exchange pulse $V_\mathrm{\varepsilon}^\mathrm{ex}$ and pulse duration $\tau_\mathrm{s}$ with the measurement point M fixed in the (0,2) charge state.
(d) $P_\mathrm{S}$ as a function of $\tau_\mathrm{s}$, extracted from the data in (b) at three different values of 
$V_\mathrm{\varepsilon}^\mathrm{ex}$. Each trace is offset by 0.7 for clarity.  Solid curves are fits to the product of a cosine and a gaussian~\cite{Petersson:2010p246804}, with amplitude, frequency, phase and decay time as free parameters. 
(e) Bloch spheres showing rotations around the axes corresponding to each trace in (c).  The angle $\theta$ between the rotation axis and $z$-axis is labeled for each case.}
\end{center}
\end{figure*}

Fig.\ \ref{fig3}(a) shows oscillations around the $z$-axis of the Bloch sphere, obtained by applying the exchange pulse sequence pioneered in~\cite{Petta:2005p2180}. Starting from point M in $S$(0,2), we first ramp from M to N at a rate that ensures fast passage through the $S$(0,2)-$T_{-}$ anticrossing, converting the state to $S$(1,1), and then ramp adiabatically from N to P, which initializes to the ground state  in the $J<h$ region.
The pulse from P to E increases $J$ suddenly so that it is comparable to or bigger than $h$, so that the rotation axis is close to the $z$-axis of the Bloch sphere.
Readout is performed by reversing the ramps, which projects
$\left |\downarrow\uparrow\right\rangle$ into the $S$(2,0) state, 
enabling readout. 
Fig.\ \ref{fig3}(c) shows the singlet probability $P_{S}$ as a function of $\tau_{S}$ and the detuning
of the exchange pulse $V_\mathrm{\varepsilon}^\mathrm{ex}$ (point E in Fig.\ 3(b)) in a range of $\varepsilon$ where $J \simgt h$.
As $V_\mathrm{\varepsilon}^\mathrm{ex}$ decreases, the oscillation frequency increases, because $J$ is increasing. The oscillation visibility also increases as the rotation axis moves towards the $z$-axis, as shown in Fig.\ \ref{fig3}(d,e).  
The inhomogeneous dephasing time $T_{2}^{*}$, extracted by fitting the time-dependence of
$P_S$ in Fig.\ \ref{fig3}(d) to
the product of a gaussian and a cosine function, 
decreases
as $J$ increases, which we argue is evidence that charge noise is limiting coherence in this regime
(see below and Fig.\ 4).

We also implemented both the $\Delta B$ and exchange gate sequences after performing
a different cycling of the external magnetic field, which resulted in a different value of
$\Delta B$, corresponding to $h\simeq32$~neV.
The results obtained are qualitatively
consistent with those shown in Figs.\ 2 and~3 (data shown in the SI Appendix, Fig. S3).

We now present evidence that the inhomogeneous dephasing is dominated by detuning noise and
by fluctuating nuclear fields, and that it does not depend on the field from the micromagnet.
Following~\cite{Dial:2013p146804}, we write
$1/T_{2}^{*}=\sqrt{\langle(\delta E_\mathrm{tot})^2\rangle}/(\sqrt{2}\hbar)$, where
$\delta E_\mathrm{tot}=\delta J (\partial E_\mathrm{tot}/\partial J) + \delta h(\partial E_\mathrm{tot}/\partial h)$, 
with
$\delta E_\mathrm{tot}$ the fluctuation in $E_\mathrm{tot}$,
$\delta J$ the fluctuation in $J$, and $\delta h$ the fluctuation in $h$.
We assume that the fluctuations in $h$ and $J$ are uncorrelated.  If fluctuations in $J$ are dominated by fluctuations in the detuning, $\delta \varepsilon$,
then $\delta J \approx \delta\varepsilon (dJ/d\varepsilon)$, and
if fluctuations in $h$ are dominated by nuclear fields, then $\delta h$ is
independent of $\varepsilon$, leading to

\begin{equation}
\sqrt{2}\hbar {T_{2}^{*}}^{-1} = \left ( \left ( \frac{J}{E_\mathrm{tot}}\frac{d J}{d \varepsilon}\delta\varepsilon_\mathrm{rms} \right )^2
+\left ( \frac{h}{E_\mathrm{tot}} \delta h_\mathrm{rms} \right )^2 \right )^{\frac{1}{2}}~,
\end{equation}
with $\delta \varepsilon_\mathrm{rms}$ and $\delta h_\mathrm{rms}$ both independent of $E_{tot}$ as well as $h$.
We use the measured $E_\mathrm{tot}$ versus $\varepsilon$ to extract $J(\varepsilon)$, which
is well-described by an exponential, $J(\varepsilon) \simeq J_{0}$exp($-\varepsilon / \varepsilon_{0}$), consistent with Ref.\ \cite{Dial:2013p146804} in the same regime {\color{black}(see SI Appendix).} 
In Fig.\ 4 we fit $T_{2}^*$ using the experimentally determined $dJ/d\varepsilon$, the measured
$E_\mathrm{tot}$, and constant values $\delta \varepsilon_\mathrm{rms} = 6.4\pm0.1~\mu$eV and $\delta h_\mathrm{rms} = 4.2\pm0.1$~neV.
The fit is good, and the values of $\delta\varepsilon_\mathrm{rms}$ and $\delta h_\mathrm{rms}$ agree well with previous reports of charge noise and fluctuations in the nuclear field in similar devices and materials~\cite{Petersson:2010p246804,Maune:2012p344,Shi:2013p075416,Taylor:2007p464,Assali:2011p165301,Culcer:2013p232108}.
The inset to Fig.\ 4, which shows data obtained at a larger $h$, demonstrates that
$T_2^*$ is well-described by Eq.\ (2) with the same $\delta\varepsilon_\mathrm{rms}$ and $\delta h_\mathrm{rms}$,
providing evidence that changing the magnetization of the micromagnet does not significantly
affect the qubit decoherence.  \mae{Equation 2 and Fig.~4 also make it clear that $T_{2}^*$ is larger for larger detunings, because charge noise has much less effect away from the primary anticrossing.}

\begin{figure}[b]
\includegraphics[width=8cm]{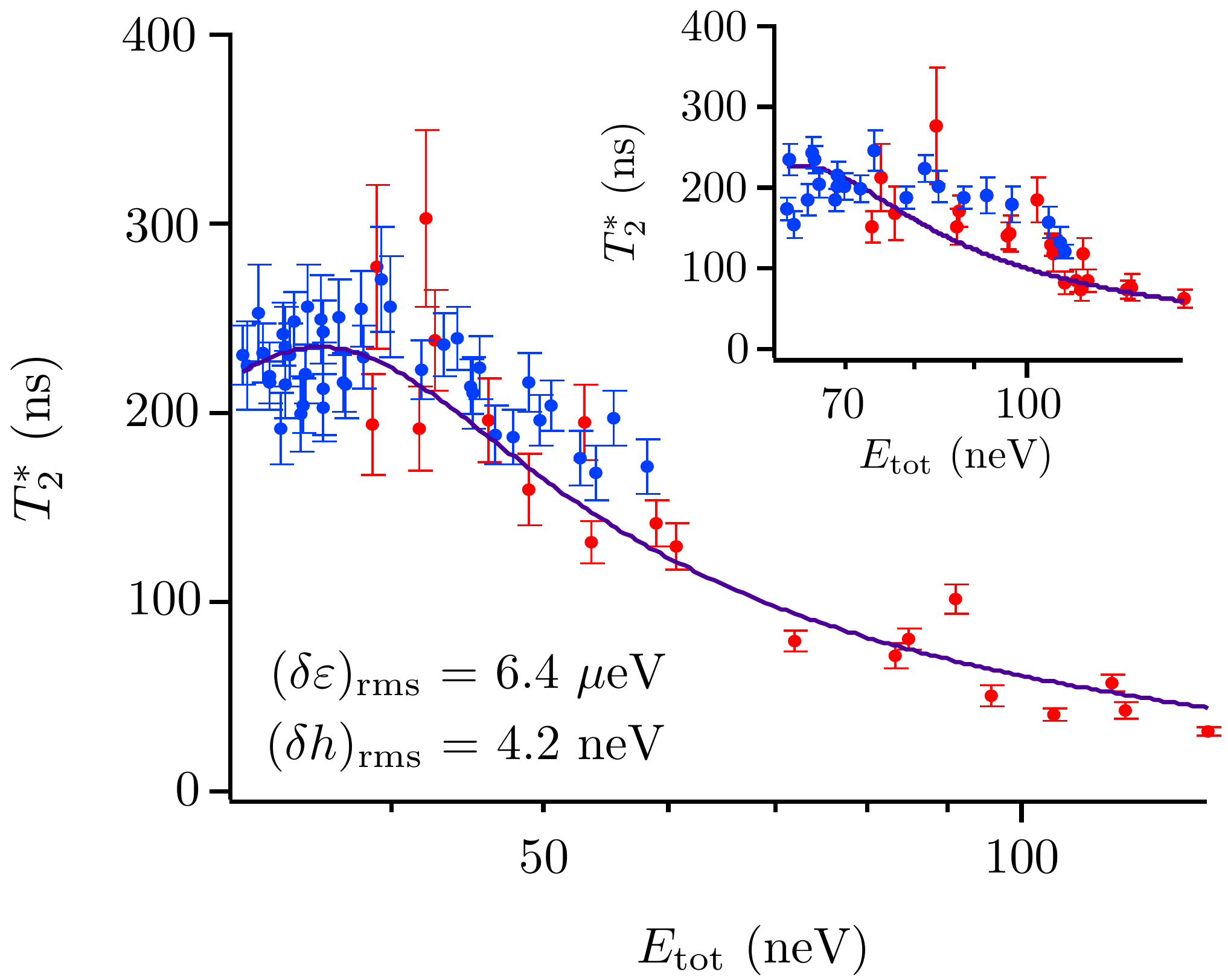}
\caption{Dependence of the inhomogeneous dephasing time $T_2^*$ on rotation energy
$E_{tot}=\sqrt{J^2+h^2}$, where $J$ is the exchange coupling and $h$ is the energy
corresponding to the magnetic field difference between the dots.  
Inset: plot of the extracted values of $T_{2}^{*}$ for 
$h\simeq$\ 60.5~neV.
Red data points are $T_{2}^{*}$ values obtained using the exchange pulse sequence (Fig.\ 3),
while blue data points are $T_{2}^{*}$ values obtained using the $\Delta B$ pulse (Fig.\ 2).
Main panel: $T_{2}^{*}$ plotted vs.\ $E_{tot}$ for $h\simeq 32$~neV, extracted from
data shown in SI Appendix Fig.~S3.
Red data points are obtained using the exchange pulse sequence (Fig.\ S3(b)),
and blue data points are obtained using the $\Delta B$ pulse (Fig.\ S3(a)).
The solid lines in the main panel and in the inset are plots of Eq.\ (2) with the
same values of $\delta \varepsilon$, the rms fluctuation in the detuning, and $\delta h$,
the rms fluctuation of the magnetic field difference, which were obtained by fitting
the data for 
$T_{2}^{*}$ as function of $E_{tot}$ at $h\simeq$~32neV to Eq.\ (2). 
The good agreement of the same form with both data sets
is strong evidence that the inhomogeneous dephasing is dominated by
charge noise and hyperfine fields and does not depend on the magnetization of
the micromagnet.
}
\label{fig4}
\end{figure}

In summary, we have demonstrated coherent  rotations of the quantum state of a singlet-triplet
qubit around two different directions of 
the Bloch sphere.
Measurements of the inhomogeneous dephasing time at a variety of exchange couplings and
two different field differences demonstrate that using an external micromagnet yields
a large increase the rotation rate about one axis on the Bloch sphere without inducing significant decoherence. 
Because the materials fabrication techniques are similar for both quantum dot-based qubits and donor-based qubits in semiconductors~\cite{Morton:2011p345}, it is reasonable to expect micromagnets also should be applicable to donor-based spin qubits~\cite{Pla:2012p489,Buch:2013p2017,Yin:2013p91}.  \mae{Micromagnets allow a difference in magnetic field to be generated between pairs of dots that does not depend on nuclear spins.  They thus offer a promising path towards fast manipulation in materials with small concentrations of nuclear spins, including both natural Si and isotopically enriched $^{28}$Si.}

{\it Acknowledgments}.  This work was supported in part by ARO (W911NF-12-0607), NSF(DMR-1206915), and by the Department of Defense.  
The views and conclusions contained in this document are those of the authors and should
not be interpreted as representing the official policies, either expressly or implied, of the US
Government.  Development
and maintenance of the growth facilities used for fabricating samples is supported by
DOE (DE-FG02-03ER46028).  This research utilized NSF-supported shared facilities at the
University of Wisconsin-Madison.

\section{Supplemental materials}

\renewcommand{\thefigure}{S\arabic{figure}}
\renewcommand{\theequation}{S\arabic{equation}}
\setcounter{figure}{0}
\setcounter{equation}{0}

{\color{black} This supplement presents methods used to calibrate the detuning energy (lever-arm $\alpha$) and to convert measurements of time-averaged current through the quantum point contact (QPC) to probabilities of being in the singlet state just after a given pulse sequence has been applied, including data used to extract the spin relaxation time $T_1$ used in the normalization process. {\color{black} Data for the ``$\Delta B$'' gate and the exchange gate for $\Delta B=32$~neV are shown here.} We present the results of a simulation of the X or ``$\Delta B$" gate performed with two different forms for the functional dependence of $J$ on detuning.  We also describe the fabrication of the sample and include an image of the micomagnet.}

{\color{black}
\subsection{Calibration of detuning energy}
We find the conversion between the detuning voltage $V_{\varepsilon}$ and the detuning energy $\varepsilon$ from measurements of the charge stability diagram under non-zero source-drain bias, as shown in Fig~S1(a). We apply $-200~\mu$V between the right dot reservoir and the left reservoir, to raise the Fermi level of the right reservoir 200$~\mu$eV higher than that of the left reservoir. By drawing the charge transition lines on top of the stability diagram, as shown in Fig.~S1(b), we can measure the shift in gate voltage of charge transitions arising from the 200~$\mu$eV potential difference between the two reservoirs. Because of the applied bias, the two triple points turn into triangles. The highlighted points are useful for converting dot energies to gate voltages. Each point has its energy level diagram drawn, as shown in Figs.~S1(c-f). Moving from the yellow point to blue point in gate voltage will raise both dot potentials by 200$~\mu$eV. Adjusting gate voltages from the yellow point (or blue point) to the green point will create a 200$~\mu$eV energy difference between the dots. The detuning direction we used is labeled with a yellow arrow, from the red point to green point, creating 200$~\mu$eV energy difference by moving each dot potential in opposite directions by the same amount. The voltage changes measured are $4$~mV on LP and $-4$~mV on RP, corresponding to $4\sqrt{2}$~mV in the detuning direction. Thus the conversion factor is $4\sqrt{2}$~mV in detuning voltage for each 200$~\mu$eV in detuning energy, corresponding to $\alpha = 35.4~\mu$eV/mV. }

\begin{figure*}
\includegraphics[width=0.7\textwidth]{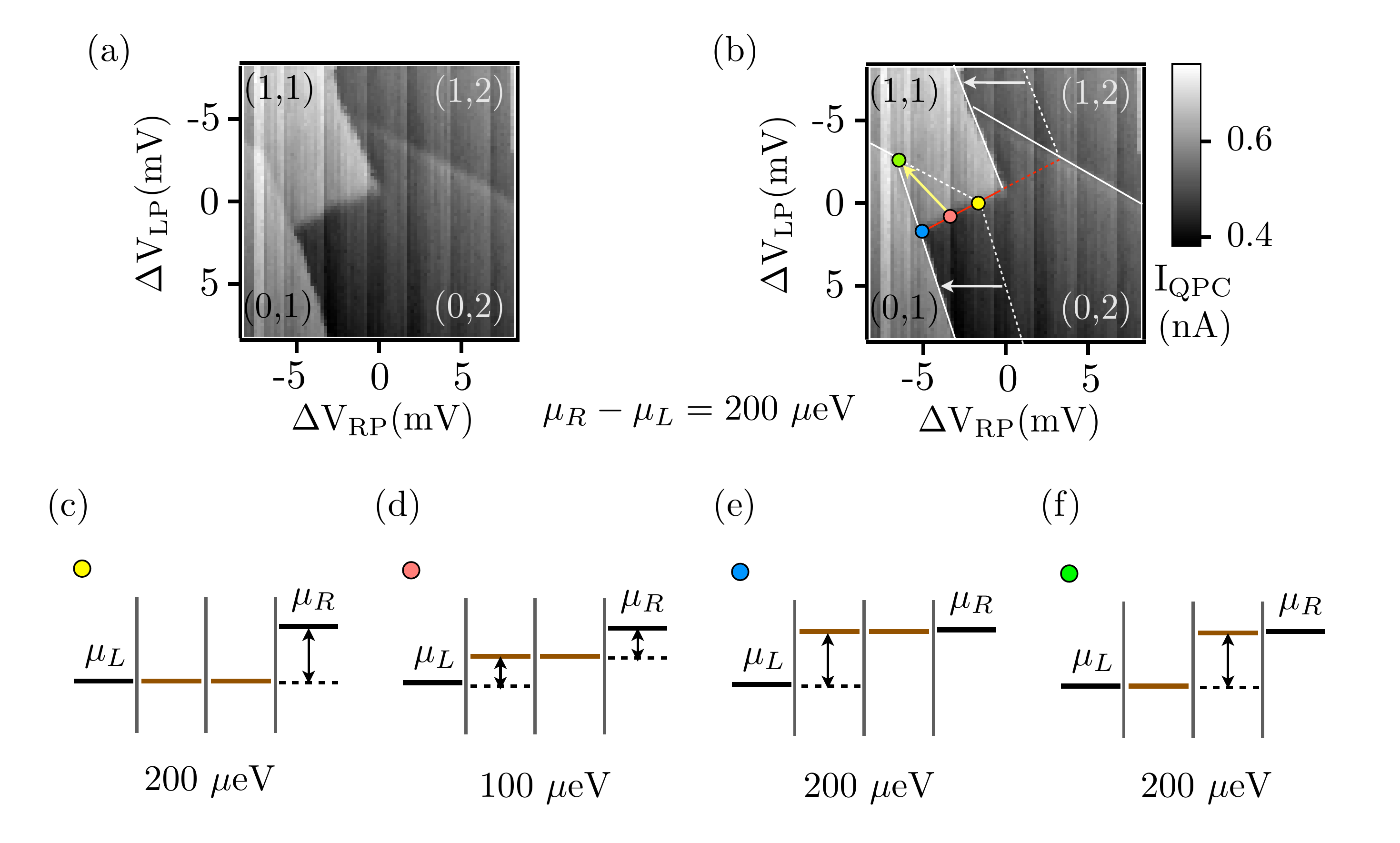}
{\color{black}\caption{(a) Charge stability diagram with -200$~\mu$V right dot reservoir bias voltage applied. Electron occupation numbers are labeled.
(b) Same charge stability diagram as (a), with charge transition lines superimposed.
A 200$~\mu$eV potential difference between left and right dot reservoirs shifts the right dot transitions $\sim-4$~mV on gate RP (see arrows). (c)-(f) Energy level diagrams showing energies for each dot and reservoir correspond to the four positions highlighted in (b). Energy differences are labeled and listed for each case. }}
\end{figure*}


\subsection{Method of conversion of the QPC current measurement to probability of being in the singlet state}
Here we present the methods used to convert measurements of the time-averaged difference in QPC current ($\Delta I_\mathrm{QPC}$) to probabilities of being in the singlet state just after a given pulse sequence has been applied. 
The method is similar to the one
described in the supplemental material of Ref.~\cite{Kim:2014p1401.4416}, except for the pulse sequence used in the extraction of the $\Delta I_\mathrm{QPC}$ that corresponds to the one electron change (0,2) to (1,1).

All the pulse sequences are generated by a Tektronix AFG3250 pulse generator. 
The reference lockin signal is a square wave with frequency of either 67 or 111~Hz (red dashed trace in Fig.~S2(a)).  During one half of a cycle, a pulse train is applied to the gates of the quantum dots (purple trace in Fig.~S2(a)). The lockin signal $\Delta I_\mathrm{QPC}$ measures the change in the average charge occupation induced by the application of the pulses. The averaging time for each data point is two seconds. 
To convert the measured $\Delta I_\mathrm{QPC}$ to singlet probability $P_\mathrm{S}$, we note
that the charge state at the end of the pulse is (1,1) for a spin triplet, while it is (0,2) for a spin singlet.
If the spin state is a triplet at the end of a pulse, it will relax back to the singlet
in a time $T_1$.
Therefore,
\begin{equation}
P_\mathrm{S} = 1-\frac{\Delta I_\mathrm{QPC}}{\Delta I_{1}}\cdot\frac{T_{1}}{T_\mathrm{m}}\cdot
\left (1-\exp\left (-\frac{T_\mathrm{m}}{T_{1}}\right )\right ),
\end{equation}
where $\Delta I_{1}$ is the value of $\Delta I_\mathrm{QPC}$ that corresponds to a one electron change from ((0,2) to (1,1)), and $T_{1}$ is the relaxation time of $T(1,1)$ to $S(0,2)$.
We measure $\Delta I_{1}$ by sweeping gate voltage along the detuning direction while applying the pulses shown in Fig.~S2(a). Fig.~S2(b) shows the lockin response as a function of detuning; the maximum change in $\Delta I_\mathrm{QPC}$ is $\Delta I_{1}$. 

The spin relaxation time $T_{1}$ for the $T_{-}$ state is extracted by measuring the $S$-$T_{-}$ oscillation amplitude as a function of $T_\mathrm{m}$, the time between successive pulses in the pulse train.  Three traces of $S$-$T_{-}$ oscillations are shown in Fig.~S2(c); they demonstrate that the oscillation amplitude decays with increasing $T_\mathrm{m}$, as expected.  The oscillation amplitude as a function of $T_\mathrm{m}$ satisfies
\begin{equation}
\Delta I_\mathrm{QPC} = A\cdot\frac{T_{1}}{T_\mathrm{m}}\cdot\left(1-\exp\left(-\frac{T_\mathrm{m}}{T_{1}}\right)\right),
\end{equation}
where $A$ is a time-independent coefficient. Fig.~S2(d) shows the oscillation amplitude as
a function of $T_{\mathrm{m}}$; a fit to Eq.~(S2) yields  $T_{1} = 9.85\pm1.19$~$\mu$s. 

To measure the spin relaxation time
$T_{1}$ for the $T_{0}$ state, we measure as a function of $T_\mathrm{m}$ the value of $\Delta I_\mathrm{QPC}$ when we pulse
into (1,1) for a time $\tau_\mathrm{s}$ significantly longer than the singlet-triplet $T_2^*$, so that  $S$ and $T_{0}$ are completely mixed ($\tau_\mathrm{s} > 2T_{2}^{*}$). 
The relaxation time for the $T_0$ state again obeys Eq.~(S2).
Fig.~S1(e) and (f) show measurements of $\Delta I_\mathrm{QPC}$ a a function of $T_\mathrm{m}$ along with
the fit to Eq.~(S2) used to
extract this $T_{1}$.


\begin{figure*}
\includegraphics[width=\textwidth]{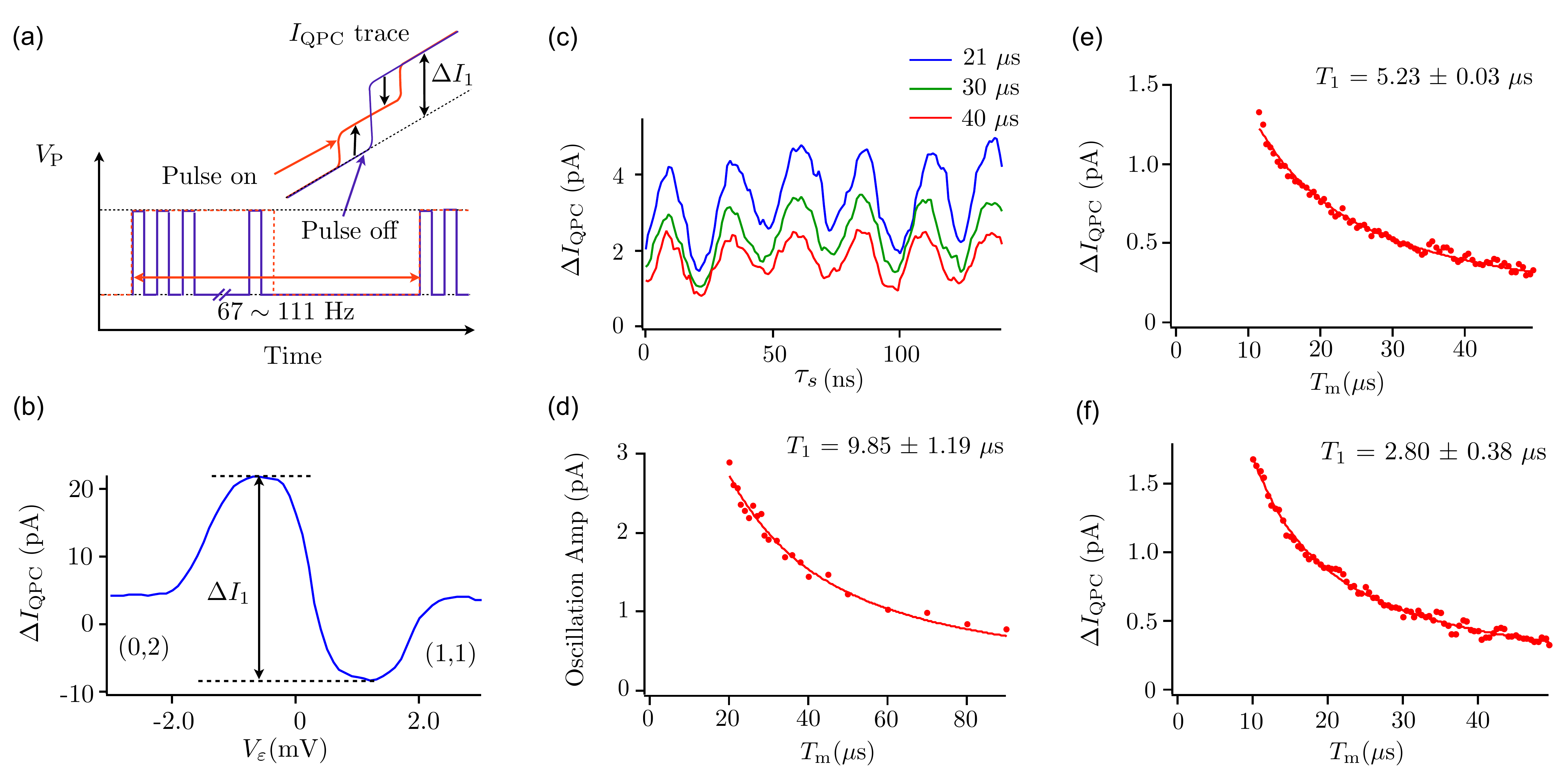}
\caption{Measurements used to determine the relationship between the lockin signal $\Delta I_\mathrm{QPC}$
and the probability of being in the singlet state after application of a given pulse sequence, using Eq.~(S1).
(a) Pulse sequence used to measure $\Delta I_\mathrm{QPC}$ that corresponds to one electron change from (0,2) to (1,1). The red dashed line indicates the low frequency signal used as the lockin reference. The square pulses shown in purple inside the red dashed line have the same frequency as the actual manipulation pulse used in the experiment. Inset: schematic diagram showing the expected dc value of $I_\mathrm{QPC}$ during the 1/2 cycle with pulses applied (red) and the 1/2 cycle with pulses not applied (purple). The black arrow indicates the maximum measured lockin signal $\Delta I_1$ (see panel (b)). 
(b) $\Delta I_\mathrm{QPC}$ measured as we sweep gate voltage along detuning with the above pulse sequence applied.
(c) Three traces of $\Delta I_\mathrm{QPC}$ as a function of pulse width $\tau_s$ are shown, exhibiting  $S$-$T_{-}$ oscillations.  The three traces are acquired with three different values of the time $T_\mathrm{m}$ between successive pulses. The oscillation amplitude decreases significantly with $T_\mathrm{m}$, indicating that relaxation to (0,2) occurs on a time
scale shorter than 20~$\mu$s.  
(d) $S$-$T_{-}$ oscillation amplitude plotted as a function of $T_\mathrm{m}$.  The corresponding value of $T_1$  is used to normalize the data shown in Fig.~1(h) in the main text. 
The solid line is a fit to Eq.~(S2), which yields the relaxation time $T_{1}$ shown on the figure. 
(e),(f) Measurement of the spin relaxation time $T_1$ for the $T_0$ state.
For this measurement, a pulse is applied into (1,1) that is significantly longer than the inhomogeneous
dephasing time $T_2^*$, so that the state at the end of the pulse is an equal mixture of $S$-$T_0$.
The decay of $\Delta I_\mathrm{QPC}$ with $T_\mathrm{m}$, the time between successive
pulses, obeys Eq.~(S2).  The value of $T_{1}$ extracted by fitting to Eq.~(S2) is listed on the plot. $T_{1}$ from (e),(f) are used to convert $\Delta I_\mathrm{QPC}$ to singlet probability for Fig.~2(c) and Fig.~3(c), respectively, in the main text.}
\end{figure*}

\subsection{Data for smaller $\Delta B$ than in main text}
{\color{black} Fig.\ S3 reports data showing oscillations in the singlet probability corresponding to the ``$\Delta B$'' gate and the exchange gate for $\Delta B=32$~neV. The pulse sequences used here are the same as those shown in Fig.~2(b) and Fig.~3(a) in the main text. }
\begin{figure*}
\includegraphics[width=0.8\textwidth]{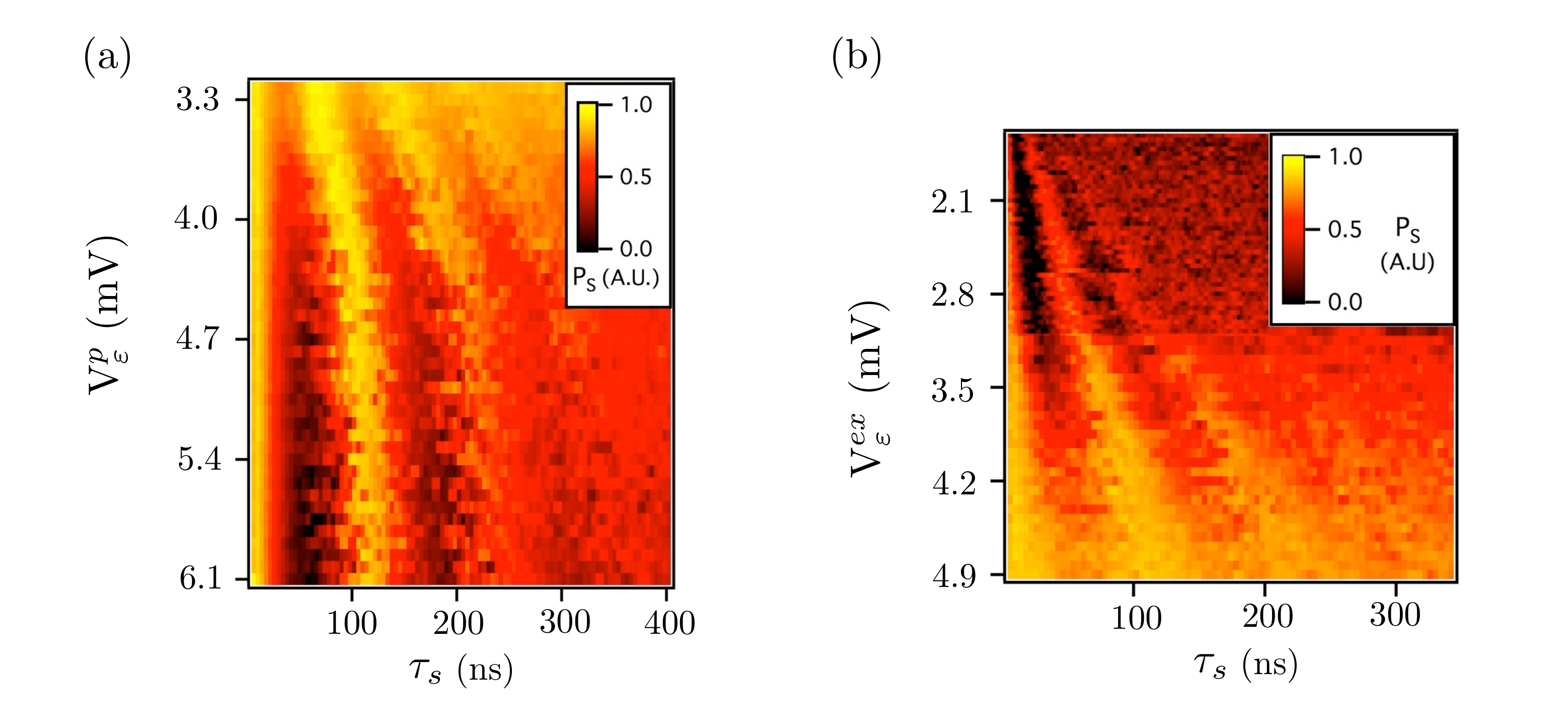}
{\color{black}\caption{Data corresponding to the ``$\Delta B$'' gate and the exchange gate for $\Delta B=32$~neV. (a) Singlet probability $P_{S}$, measured as a function of pulse duration and voltage level at pulse tip, $V_\varepsilon^{p}$. (b) Singlet probability  $P_{S}$, measured as a function of pulse duration and pulse level $V_\varepsilon^{ex}$ in the exchange pulse sequence. The singlet probability is reported in arbitrary units in both (a) and (b). }}
\end{figure*}

\subsection{Simulation of the X or ``$\Delta$B'' gate}
Fig.\ S4 reports the results simulations of the ``$\Delta$B'' gate with two different functional forms for the dependence of $J$ on detuning, with the details described in the caption.  We find that $J$ appears to vary exponentially as a function of detuning energy, in agreement with previous observations by Dial et al.~\cite{Dial:2013p146804}.
\vspace{.5cm}

\subsection{Micromagnet fabrication}

An optical micrograph of the device including the micromagnet is shown in Fig.\ S5.  The micromagnet is 12.64~$\mu$m $\times$ 1.78~$\mu$m $\times$~242 nm.  The magnet was patterned via electron-beam lithography on top of the accumulation gates approximately 1.78~$\mu$m to the left and 122 nm above the center of the two quantum dots.  The magnet was deposited via electron-beam evaporation with a metal film stack of 2~nm Ti / 20~nm Au / 200~nm Co / 20 ~nm Au evaporated at approximately 0.3 \AA/s.  The gold film is intended to help minimize oxidation of the Co film.  

\begin{figure*}
\includegraphics[width=0.8\textwidth]{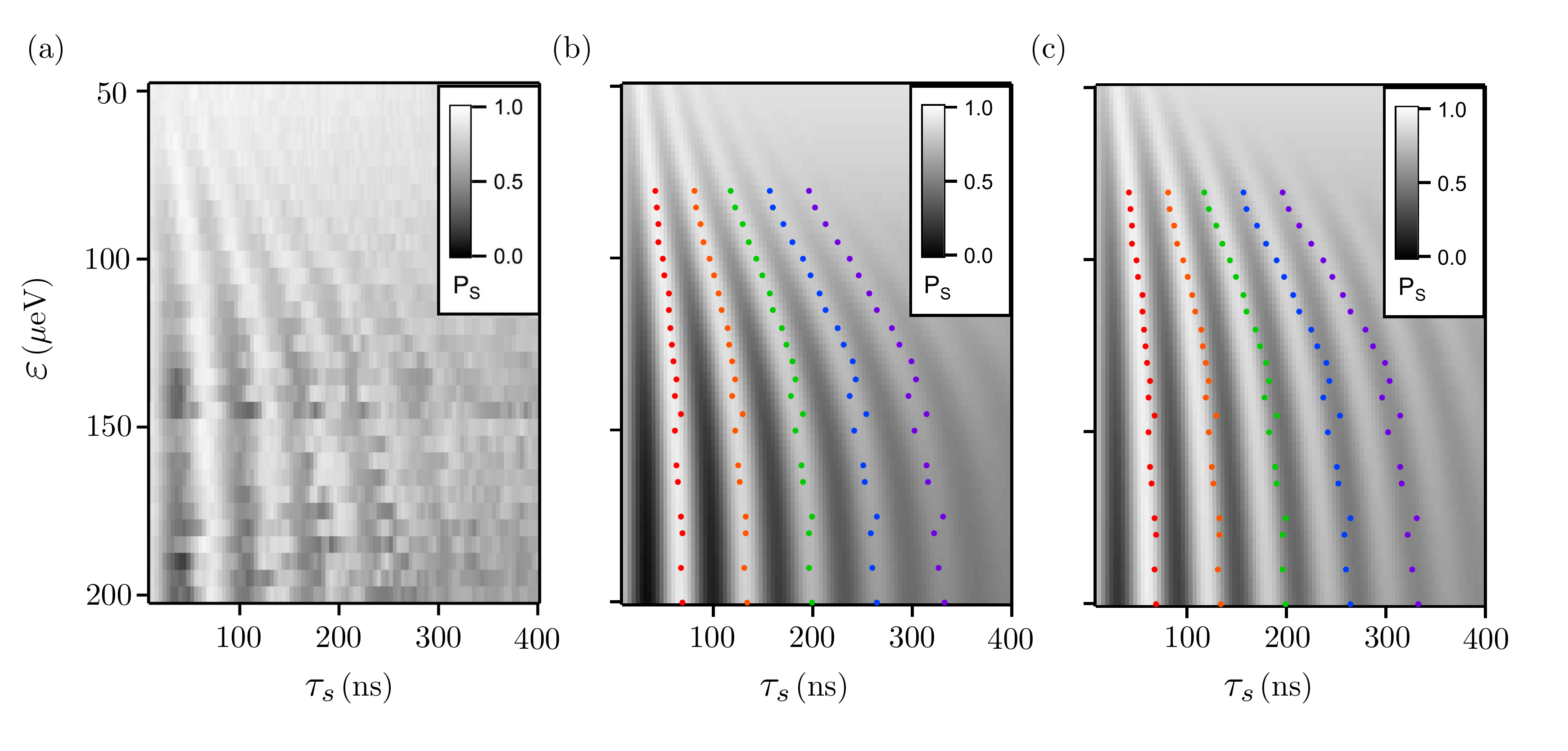}
{\color{black}\caption{``$\Delta B$'' gate data compared to simulation results using different functional forms for the dependence of $J$ on detuning. (a) Experimentally measured singlet probability $P_\mathrm{S}$ plotted as a function of pulse duration $\tau_{s}$ and detuning energy $\varepsilon$ at the pulse tip (Fig.~2(c) from the main text).  Line cuts of data are fit to products of sinusoids and Gaussians, and the resulting maxima from the fits are plotted as colored dots in panels (b) and (c).  (b) Simulation of singlet probability $P_\mathrm{S}$ as function of duration $\tau_{s}$ and detuning energy $\varepsilon$ using $J=J_{0}\exp(-\varepsilon/\varepsilon_{0})$, where the best fit is found with $\varepsilon_{0} = 62.7~\mu eV$. (c) Simulation of singlet probability $P_\mathrm{S}$ as function of duration $\tau_{s}$ and detuning energy $\varepsilon$ using $J=\sqrt{\varepsilon^{2}/4+t_{c}^{2}}-\varepsilon/2$, where the best fit is found with $t_{c} = 2.48~\mu eV$. Oscillation peaks extracted from (a) are plotted on top of (b) and (c), and the comparison suggests that $J=J_{0}\exp(-\varepsilon/\varepsilon_{0})$ fits the data well. }}
\end{figure*}

\begin{figure}
\includegraphics[width=0.4\textwidth]{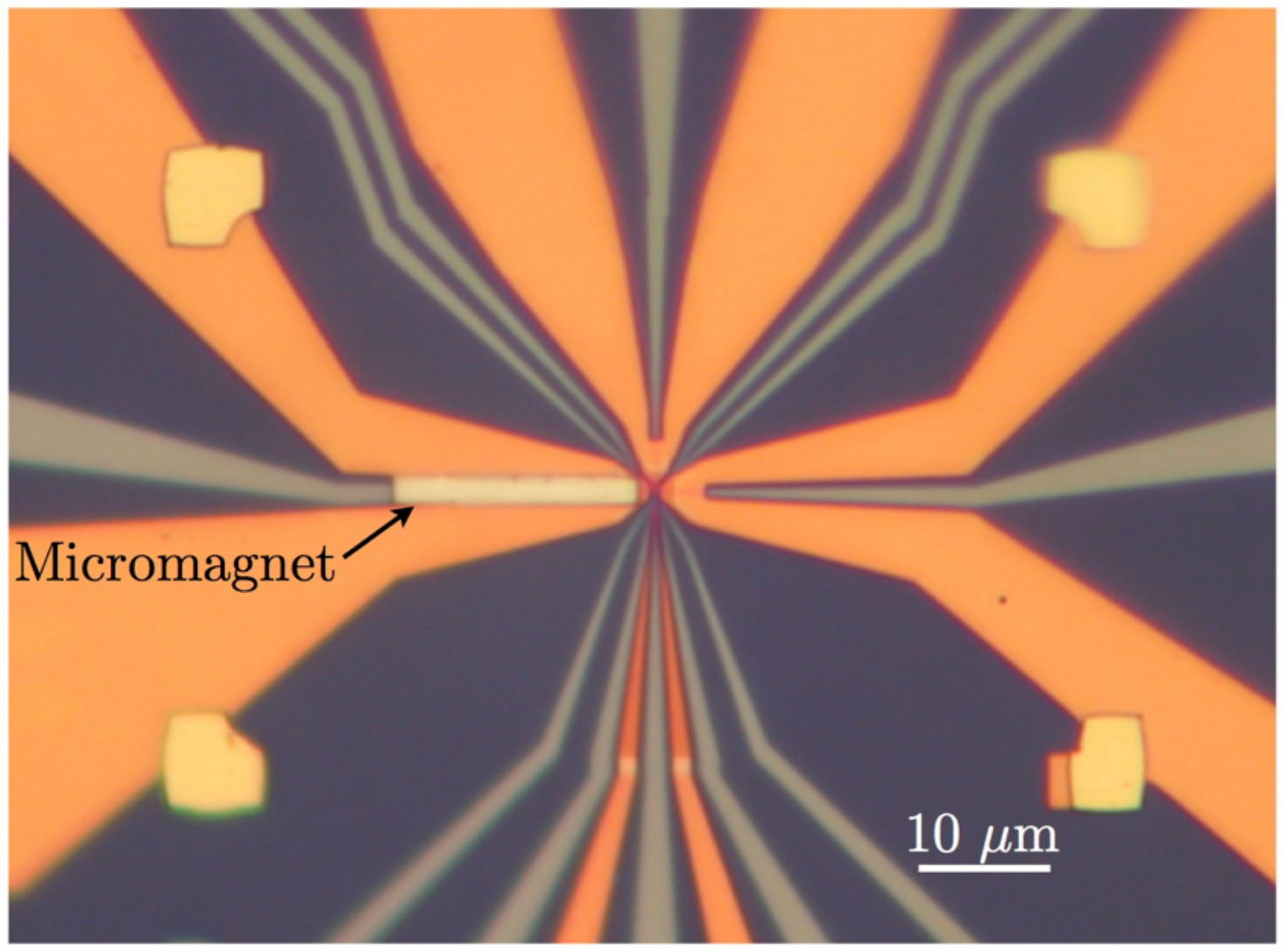}
\caption{Optical micrograph of the device, with the location of the micromagnet marked on the figure.}
\end{figure}

\bibliography{siliconqcsnc}

\begin{thebibliography}{38}%
\makeatletter
\providecommand \@ifxundefined [1]{%
 \@ifx{#1\undefined}
}%
\providecommand \@ifnum [1]{%
 \ifnum #1\expandafter \@firstoftwo
 \else \expandafter \@secondoftwo
 \fi
}%
\providecommand \@ifx [1]{%
 \ifx #1\expandafter \@firstoftwo
 \else \expandafter \@secondoftwo
 \fi
}%
\providecommand \natexlab [1]{#1}%
\providecommand \enquote  [1]{``#1''}%
\providecommand \bibnamefont  [1]{#1}%
\providecommand \bibfnamefont [1]{#1}%
\providecommand \citenamefont [1]{#1}%
\providecommand \href@noop [0]{\@secondoftwo}%
\providecommand \href [0]{\begingroup \@sanitize@url \@href}%
\providecommand \@href[1]{\@@startlink{#1}\@@href}%
\providecommand \@@href[1]{\endgroup#1\@@endlink}%
\providecommand \@sanitize@url [0]{\catcode `\\12\catcode `\$12\catcode
  `\&12\catcode `\#12\catcode `\^12\catcode `\_12\catcode `\%12\relax}%
\providecommand \@@startlink[1]{}%
\providecommand \@@endlink[0]{}%
\providecommand \url  [0]{\begingroup\@sanitize@url \@url }%
\providecommand \@url [1]{\endgroup\@href {#1}{\urlprefix }}%
\providecommand \urlprefix  [0]{URL }%
\providecommand \Eprint [0]{\href }%
\providecommand \doibase [0]{http://dx.doi.org/}%
\providecommand \selectlanguage [0]{\@gobble}%
\providecommand \bibinfo  [0]{\@secondoftwo}%
\providecommand \bibfield  [0]{\@secondoftwo}%
\providecommand \translation [1]{[#1]}%
\providecommand \BibitemOpen [0]{}%
\providecommand \bibitemStop [0]{}%
\providecommand \bibitemNoStop [0]{.\EOS\space}%
\providecommand \EOS [0]{\spacefactor3000\relax}%
\providecommand \BibitemShut  [1]{\csname bibitem#1\endcsname}%
\let\auto@bib@innerbib\@empty
\bibitem [{\citenamefont {Levy}(2002)}]{Levy:2002p1446}%
  \BibitemOpen
  \bibfield  {author} {\bibinfo {author} {\bibfnamefont {J.}~\bibnamefont
  {Levy}},\ }\href {\doibase 10.1103/PhysRevLett.89.147902} {\bibfield
  {journal} {\bibinfo  {journal} {Phys. Rev. Lett.}\ }\textbf {\bibinfo
  {volume} {89}},\ \bibinfo {pages} {147902} (\bibinfo {year}
  {2002})}\BibitemShut {NoStop}%
\bibitem [{\citenamefont {Petta}\ \emph {et~al.}(2005)\citenamefont {Petta},
  \citenamefont {Johnson}, \citenamefont {Taylor}, \citenamefont {Laird},
  \citenamefont {Yacoby}, \citenamefont {Lukin}, \citenamefont {Marcus},
  \citenamefont {Hanson},\ and\ \citenamefont {Gossard}}]{Petta:2005p2180}%
  \BibitemOpen
  \bibfield  {author} {\bibinfo {author} {\bibfnamefont {J.~R.}\ \bibnamefont
  {Petta}}, \bibinfo {author} {\bibfnamefont {A.~C.}\ \bibnamefont {Johnson}},
  \bibinfo {author} {\bibfnamefont {J.~M.}\ \bibnamefont {Taylor}}, \bibinfo
  {author} {\bibfnamefont {E.~A.}\ \bibnamefont {Laird}}, \bibinfo {author}
  {\bibfnamefont {A.}~\bibnamefont {Yacoby}}, \bibinfo {author} {\bibfnamefont
  {M.~D.}\ \bibnamefont {Lukin}}, \bibinfo {author} {\bibfnamefont {C.~M.}\
  \bibnamefont {Marcus}}, \bibinfo {author} {\bibfnamefont {M.~P.}\
  \bibnamefont {Hanson}}, \ and\ \bibinfo {author} {\bibfnamefont {A.~C.}\
  \bibnamefont {Gossard}},\ }\href {\doibase 10.1126/science.1116955}
  {\bibfield  {journal} {\bibinfo  {journal} {Science}\ }\textbf {\bibinfo
  {volume} {309}},\ \bibinfo {pages} {2180} (\bibinfo {year}
  {2005})}\BibitemShut {NoStop}%
\bibitem [{\citenamefont {Nowack}\ \emph {et~al.}(2011)\citenamefont {Nowack},
  \citenamefont {Shafiei}, \citenamefont {Laforest}, \citenamefont
  {Prawiroatmodjo}, \citenamefont {Schreiber}, \citenamefont {Reichl},
  \citenamefont {Wegscheider},\ and\ \citenamefont
  {Vandersypen}}]{Nowack:2011p1269}%
  \BibitemOpen
  \bibfield  {author} {\bibinfo {author} {\bibfnamefont {K.}~\bibnamefont
  {Nowack}}, \bibinfo {author} {\bibfnamefont {M.}~\bibnamefont {Shafiei}},
  \bibinfo {author} {\bibfnamefont {M.}~\bibnamefont {Laforest}}, \bibinfo
  {author} {\bibfnamefont {G.}~\bibnamefont {Prawiroatmodjo}}, \bibinfo
  {author} {\bibfnamefont {L.}~\bibnamefont {Schreiber}}, \bibinfo {author}
  {\bibfnamefont {C.}~\bibnamefont {Reichl}}, \bibinfo {author} {\bibfnamefont
  {W.}~\bibnamefont {Wegscheider}}, \ and\ \bibinfo {author} {\bibfnamefont
  {L.}~\bibnamefont {Vandersypen}},\ }\href@noop {} {\bibfield  {journal}
  {\bibinfo  {journal} {Science}\ }\textbf {\bibinfo {volume} {333}},\ \bibinfo
  {pages} {1269} (\bibinfo {year} {2011})}\BibitemShut {NoStop}%
\bibitem [{\citenamefont {Shi}\ \emph {et~al.}(2014)\citenamefont {Shi},
  \citenamefont {Simmons}, \citenamefont {Ward}, \citenamefont {Prance},
  \citenamefont {Wu}, \citenamefont {Koh}, \citenamefont {Gamble},
  \citenamefont {Savage}, \citenamefont {Lagally}, \citenamefont {Friesen},
  \citenamefont {Coppersmith},\ and\ \citenamefont {Eriksson}}]{Shi:2014p3020}%
  \BibitemOpen
  \bibfield  {author} {\bibinfo {author} {\bibfnamefont {Z.}~\bibnamefont
  {Shi}}, \bibinfo {author} {\bibfnamefont {C.~B.}\ \bibnamefont {Simmons}},
  \bibinfo {author} {\bibfnamefont {D.~R.}\ \bibnamefont {Ward}}, \bibinfo
  {author} {\bibfnamefont {J.~R.}\ \bibnamefont {Prance}}, \bibinfo {author}
  {\bibfnamefont {X.}~\bibnamefont {Wu}}, \bibinfo {author} {\bibfnamefont
  {T.~S.}\ \bibnamefont {Koh}}, \bibinfo {author} {\bibfnamefont {J.~K.}\
  \bibnamefont {Gamble}}, \bibinfo {author} {\bibfnamefont {D.~E.}\
  \bibnamefont {Savage}}, \bibinfo {author} {\bibfnamefont {M.~G.}\
  \bibnamefont {Lagally}}, \bibinfo {author} {\bibfnamefont {M.}~\bibnamefont
  {Friesen}}, \bibinfo {author} {\bibfnamefont {S.~N.}\ \bibnamefont
  {Coppersmith}}, \ and\ \bibinfo {author} {\bibfnamefont {M.~A.}\ \bibnamefont
  {Eriksson}},\ }\href@noop {} {\bibfield  {journal} {\bibinfo  {journal}
  {Nature Comm.}\ }\textbf {\bibinfo {volume} {5}},\ \bibinfo {pages} {3020}
  (\bibinfo {year} {2014})}\BibitemShut {NoStop}%
\bibitem [{\citenamefont {Kim}\ \emph {et~al.}(2014)\citenamefont {Kim},
  \citenamefont {Shi}, \citenamefont {Simmons}, \citenamefont {Ward},
  \citenamefont {Prance}, \citenamefont {Koh}, \citenamefont {Gamble},
  \citenamefont {Savage}, \citenamefont {Lagally}, \citenamefont {Friesen},
  \citenamefont {Coppersmith},\ and\ \citenamefont
  {Eriksson}}]{Kim:2014p1401.4416}%
  \BibitemOpen
  \bibfield  {author} {\bibinfo {author} {\bibfnamefont {D.}~\bibnamefont
  {Kim}}, \bibinfo {author} {\bibfnamefont {Z.}~\bibnamefont {Shi}}, \bibinfo
  {author} {\bibfnamefont {C.~B.}\ \bibnamefont {Simmons}}, \bibinfo {author}
  {\bibfnamefont {D.~R.}\ \bibnamefont {Ward}}, \bibinfo {author}
  {\bibfnamefont {J.~R.}\ \bibnamefont {Prance}}, \bibinfo {author}
  {\bibfnamefont {T.~S.}\ \bibnamefont {Koh}}, \bibinfo {author} {\bibfnamefont
  {J.~K.}\ \bibnamefont {Gamble}}, \bibinfo {author} {\bibfnamefont {D.~E.}\
  \bibnamefont {Savage}}, \bibinfo {author} {\bibfnamefont {M.~G.}\
  \bibnamefont {Lagally}}, \bibinfo {author} {\bibfnamefont {M.}~\bibnamefont
  {Friesen}}, \bibinfo {author} {\bibfnamefont {S.~N.}\ \bibnamefont
  {Coppersmith}}, \ and\ \bibinfo {author} {\bibfnamefont {M.~A.}\ \bibnamefont
  {Eriksson}},\ }\href@noop {} {\enquote {\bibinfo {title} {Quantum control and
  process tomography of a semiconductor quantum dot hybrid qubit},}\ }
  (\bibinfo {year} {2014}),\ \bibinfo {note} {preprint
  arXiv:1401.4416}\BibitemShut {NoStop}%
\bibitem [{\citenamefont {Gaudreau}\ \emph {et~al.}(2012)\citenamefont
  {Gaudreau}, \citenamefont {Granger}, \citenamefont {Kam}, \citenamefont
  {Aers}, \citenamefont {Studenikin}, \citenamefont {Zawadzki}, \citenamefont
  {Pioro-Ladri\`{e}re}, \citenamefont {Wasilewski},\ and\ \citenamefont
  {Sachrajda}}]{Gaudreau:2011p54}%
  \BibitemOpen
  \bibfield  {author} {\bibinfo {author} {\bibfnamefont {L.}~\bibnamefont
  {Gaudreau}}, \bibinfo {author} {\bibfnamefont {G.}~\bibnamefont {Granger}},
  \bibinfo {author} {\bibfnamefont {A.}~\bibnamefont {Kam}}, \bibinfo {author}
  {\bibfnamefont {G.~C.}\ \bibnamefont {Aers}}, \bibinfo {author}
  {\bibfnamefont {S.~A.}\ \bibnamefont {Studenikin}}, \bibinfo {author}
  {\bibfnamefont {P.}~\bibnamefont {Zawadzki}}, \bibinfo {author}
  {\bibfnamefont {M.}~\bibnamefont {Pioro-Ladri\`{e}re}}, \bibinfo {author}
  {\bibfnamefont {Z.~R.}\ \bibnamefont {Wasilewski}}, \ and\ \bibinfo {author}
  {\bibfnamefont {A.~S.}\ \bibnamefont {Sachrajda}},\ }\href@noop {} {\bibfield
   {journal} {\bibinfo  {journal} {Nature Physics}\ }\textbf {\bibinfo {volume}
  {8}},\ \bibinfo {pages} {54} (\bibinfo {year} {2012})}\BibitemShut {NoStop}%
\bibitem [{\citenamefont {Medford}\ \emph {et~al.}(2013)\citenamefont
  {Medford}, \citenamefont {Beil}, \citenamefont {Taylor}, \citenamefont
  {Bartlett}, \citenamefont {Doherty}, \citenamefont {Rashba}, \citenamefont
  {DiVincenzo}, \citenamefont {Lu}, \citenamefont {Gossard},\ and\
  \citenamefont {Marcus}}]{Medford:2013p654}%
  \BibitemOpen
  \bibfield  {author} {\bibinfo {author} {\bibfnamefont {J.}~\bibnamefont
  {Medford}}, \bibinfo {author} {\bibfnamefont {J.}~\bibnamefont {Beil}},
  \bibinfo {author} {\bibfnamefont {J.~M.}\ \bibnamefont {Taylor}}, \bibinfo
  {author} {\bibfnamefont {S.~D.}\ \bibnamefont {Bartlett}}, \bibinfo {author}
  {\bibfnamefont {A.~C.}\ \bibnamefont {Doherty}}, \bibinfo {author}
  {\bibfnamefont {E.~I.}\ \bibnamefont {Rashba}}, \bibinfo {author}
  {\bibfnamefont {D.~P.}\ \bibnamefont {DiVincenzo}}, \bibinfo {author}
  {\bibfnamefont {H.}~\bibnamefont {Lu}}, \bibinfo {author} {\bibfnamefont
  {A.~C.}\ \bibnamefont {Gossard}}, \ and\ \bibinfo {author} {\bibfnamefont
  {C.~M.}\ \bibnamefont {Marcus}},\ }\href@noop {} {\bibfield  {journal}
  {\bibinfo  {journal} {Nat. Nano.}\ }\textbf {\bibinfo {volume} {8}},\
  \bibinfo {pages} {654} (\bibinfo {year} {2013})}\BibitemShut {NoStop}%
\bibitem [{\citenamefont {Shulman}\ \emph {et~al.}(2012)\citenamefont
  {Shulman}, \citenamefont {Dial}, \citenamefont {Harvey}, \citenamefont
  {Bluhm}, \citenamefont {Umansky},\ and\ \citenamefont
  {Yacoby}}]{Shulman:2012p202}%
  \BibitemOpen
  \bibfield  {author} {\bibinfo {author} {\bibfnamefont {M.~D.}\ \bibnamefont
  {Shulman}}, \bibinfo {author} {\bibfnamefont {O.~E.}\ \bibnamefont {Dial}},
  \bibinfo {author} {\bibfnamefont {S.~P.}\ \bibnamefont {Harvey}}, \bibinfo
  {author} {\bibfnamefont {H.}~\bibnamefont {Bluhm}}, \bibinfo {author}
  {\bibfnamefont {V.}~\bibnamefont {Umansky}}, \ and\ \bibinfo {author}
  {\bibfnamefont {A.}~\bibnamefont {Yacoby}},\ }\href@noop {} {\bibfield
  {journal} {\bibinfo  {journal} {Science}\ }\textbf {\bibinfo {volume}
  {336}},\ \bibinfo {pages} {202} (\bibinfo {year} {2012})}\BibitemShut
  {NoStop}%
\bibitem [{\citenamefont {Reilly}\ \emph {et~al.}(2008)\citenamefont {Reilly},
  \citenamefont {Taylor}, \citenamefont {Petta}, \citenamefont {Marcus},
  \citenamefont {Hanson},\ and\ \citenamefont {Gossard}}]{Reilly:2008p817}%
  \BibitemOpen
  \bibfield  {author} {\bibinfo {author} {\bibfnamefont {D.~J.}\ \bibnamefont
  {Reilly}}, \bibinfo {author} {\bibfnamefont {J.~M.}\ \bibnamefont {Taylor}},
  \bibinfo {author} {\bibfnamefont {J.~R.}\ \bibnamefont {Petta}}, \bibinfo
  {author} {\bibfnamefont {C.~M.}\ \bibnamefont {Marcus}}, \bibinfo {author}
  {\bibfnamefont {M.~P.}\ \bibnamefont {Hanson}}, \ and\ \bibinfo {author}
  {\bibfnamefont {A.~C.}\ \bibnamefont {Gossard}},\ }\href {\doibase
  10.1126/science.1159221} {\bibfield  {journal} {\bibinfo  {journal}
  {Science}\ }\textbf {\bibinfo {volume} {321}},\ \bibinfo {pages} {817}
  (\bibinfo {year} {2008})}\BibitemShut {NoStop}%
\bibitem [{\citenamefont {Foletti}\ \emph {et~al.}(2009)\citenamefont
  {Foletti}, \citenamefont {Bluhm}, \citenamefont {Mahalu}, \citenamefont
  {Umansky},\ and\ \citenamefont {Yacoby}}]{Foletti:2009p903}%
  \BibitemOpen
  \bibfield  {author} {\bibinfo {author} {\bibfnamefont {S.}~\bibnamefont
  {Foletti}}, \bibinfo {author} {\bibfnamefont {H.}~\bibnamefont {Bluhm}},
  \bibinfo {author} {\bibfnamefont {D.}~\bibnamefont {Mahalu}}, \bibinfo
  {author} {\bibfnamefont {V.}~\bibnamefont {Umansky}}, \ and\ \bibinfo
  {author} {\bibfnamefont {A.}~\bibnamefont {Yacoby}},\ }\href@noop {}
  {\bibfield  {journal} {\bibinfo  {journal} {Nature Physics}\ }\textbf
  {\bibinfo {volume} {5}},\ \bibinfo {pages} {903} (\bibinfo {year}
  {2009})}\BibitemShut {NoStop}%
\bibitem [{\citenamefont {Barthel}\ \emph {et~al.}(2009)\citenamefont
  {Barthel}, \citenamefont {Reilly}, \citenamefont {Marcus}, \citenamefont
  {Hanson},\ and\ \citenamefont {Gossard}}]{Barthel:2009p160503}%
  \BibitemOpen
  \bibfield  {author} {\bibinfo {author} {\bibfnamefont {C.}~\bibnamefont
  {Barthel}}, \bibinfo {author} {\bibfnamefont {D.~J.}\ \bibnamefont {Reilly}},
  \bibinfo {author} {\bibfnamefont {C.~M.}\ \bibnamefont {Marcus}}, \bibinfo
  {author} {\bibfnamefont {M.~P.}\ \bibnamefont {Hanson}}, \ and\ \bibinfo
  {author} {\bibfnamefont {A.~C.}\ \bibnamefont {Gossard}},\ }\href@noop {}
  {\bibfield  {journal} {\bibinfo  {journal} {Phys. Rev. Lett.}\ }\textbf
  {\bibinfo {volume} {103}},\ \bibinfo {pages} {160503} (\bibinfo {year}
  {2009})}\BibitemShut {NoStop}%
\bibitem [{\citenamefont {Barthel}\ \emph {et~al.}(2010)\citenamefont
  {Barthel}, \citenamefont {Medford}, \citenamefont {Marcus}, \citenamefont
  {Hanson},\ and\ \citenamefont {Gossard}}]{Barthel:2010p266808}%
  \BibitemOpen
  \bibfield  {author} {\bibinfo {author} {\bibfnamefont {C.}~\bibnamefont
  {Barthel}}, \bibinfo {author} {\bibfnamefont {J.}~\bibnamefont {Medford}},
  \bibinfo {author} {\bibfnamefont {C.}~\bibnamefont {Marcus}}, \bibinfo
  {author} {\bibfnamefont {M.}~\bibnamefont {Hanson}}, \ and\ \bibinfo {author}
  {\bibfnamefont {A.}~\bibnamefont {Gossard}},\ }\href@noop {} {\bibfield
  {journal} {\bibinfo  {journal} {Phys. Rev. Lett.}\ }\textbf {\bibinfo
  {volume} {105}},\ \bibinfo {pages} {266808} (\bibinfo {year}
  {2010})}\BibitemShut {NoStop}%
\bibitem [{\citenamefont {{Bluhm}}\ \emph {et~al.}(2011)\citenamefont
  {{Bluhm}}, \citenamefont {{Foletti}}, \citenamefont {{Neder}}, \citenamefont
  {{Rudner}}, \citenamefont {{Mahalu}}, \citenamefont {{Umansky}},\ and\
  \citenamefont {{Yacoby}}}]{Bluhm:2011p109}%
  \BibitemOpen
  \bibfield  {author} {\bibinfo {author} {\bibfnamefont {H.}~\bibnamefont
  {{Bluhm}}}, \bibinfo {author} {\bibfnamefont {S.}~\bibnamefont {{Foletti}}},
  \bibinfo {author} {\bibfnamefont {I.}~\bibnamefont {{Neder}}}, \bibinfo
  {author} {\bibfnamefont {M.}~\bibnamefont {{Rudner}}}, \bibinfo {author}
  {\bibfnamefont {D.}~\bibnamefont {{Mahalu}}}, \bibinfo {author}
  {\bibfnamefont {V.}~\bibnamefont {{Umansky}}}, \ and\ \bibinfo {author}
  {\bibfnamefont {A.}~\bibnamefont {{Yacoby}}},\ }\href {\doibase
  10.1038/nphys1856} {\bibfield  {journal} {\bibinfo  {journal} {Nat. Phys.}\
  }\textbf {\bibinfo {volume} {7}},\ \bibinfo {pages} {109} (\bibinfo {year}
  {2011})}\BibitemShut {NoStop}%
\bibitem [{\citenamefont {Maune}\ \emph {et~al.}(2012)\citenamefont {Maune},
  \citenamefont {Borselli}, \citenamefont {Huang}, \citenamefont {Ladd},
  \citenamefont {Deelman}, \citenamefont {Holabird}, \citenamefont {Kiselev},
  \citenamefont {Alvarado-Rodriguez}, \citenamefont {Ross}, \citenamefont
  {Schmitz}, \citenamefont {Sokolich}, \citenamefont {Watson}, \citenamefont
  {Gyure},\ and\ \citenamefont {Hunter}}]{Maune:2012p344}%
  \BibitemOpen
  \bibfield  {author} {\bibinfo {author} {\bibfnamefont {B.~M.}\ \bibnamefont
  {Maune}}, \bibinfo {author} {\bibfnamefont {M.~G.}\ \bibnamefont {Borselli}},
  \bibinfo {author} {\bibfnamefont {B.}~\bibnamefont {Huang}}, \bibinfo
  {author} {\bibfnamefont {T.~D.}\ \bibnamefont {Ladd}}, \bibinfo {author}
  {\bibfnamefont {P.~W.}\ \bibnamefont {Deelman}}, \bibinfo {author}
  {\bibfnamefont {K.~S.}\ \bibnamefont {Holabird}}, \bibinfo {author}
  {\bibfnamefont {A.~A.}\ \bibnamefont {Kiselev}}, \bibinfo {author}
  {\bibfnamefont {I.}~\bibnamefont {Alvarado-Rodriguez}}, \bibinfo {author}
  {\bibfnamefont {R.~S.}\ \bibnamefont {Ross}}, \bibinfo {author}
  {\bibfnamefont {A.~E.}\ \bibnamefont {Schmitz}}, \bibinfo {author}
  {\bibfnamefont {M.}~\bibnamefont {Sokolich}}, \bibinfo {author}
  {\bibfnamefont {C.~A.}\ \bibnamefont {Watson}}, \bibinfo {author}
  {\bibfnamefont {M.~F.}\ \bibnamefont {Gyure}}, \ and\ \bibinfo {author}
  {\bibfnamefont {A.~T.}\ \bibnamefont {Hunter}},\ }\href@noop {} {\bibfield
  {journal} {\bibinfo  {journal} {Nature}\ }\textbf {\bibinfo {volume} {481}},\
  \bibinfo {pages} {344} (\bibinfo {year} {2012})}\BibitemShut {NoStop}%
\bibitem [{\citenamefont {Shi}\ \emph {et~al.}(2011)\citenamefont {Shi},
  \citenamefont {Simmons}, \citenamefont {Prance}, \citenamefont {Gamble},
  \citenamefont {Friesen}, \citenamefont {Savage}, \citenamefont {Lagally},
  \citenamefont {Coppersmith},\ and\ \citenamefont
  {Eriksson}}]{Shi:2011p233108}%
  \BibitemOpen
  \bibfield  {author} {\bibinfo {author} {\bibfnamefont {Z.}~\bibnamefont
  {Shi}}, \bibinfo {author} {\bibfnamefont {C.~B.}\ \bibnamefont {Simmons}},
  \bibinfo {author} {\bibfnamefont {J.}~\bibnamefont {Prance}}, \bibinfo
  {author} {\bibfnamefont {J.~K.}\ \bibnamefont {Gamble}}, \bibinfo {author}
  {\bibfnamefont {M.}~\bibnamefont {Friesen}}, \bibinfo {author} {\bibfnamefont
  {D.~E.}\ \bibnamefont {Savage}}, \bibinfo {author} {\bibfnamefont {M.~G.}\
  \bibnamefont {Lagally}}, \bibinfo {author} {\bibfnamefont {S.~N.}\
  \bibnamefont {Coppersmith}}, \ and\ \bibinfo {author} {\bibfnamefont {M.~A.}\
  \bibnamefont {Eriksson}},\ }\href@noop {} {\bibfield  {journal} {\bibinfo
  {journal} {Appl. Phys. Lett.}\ }\textbf {\bibinfo {volume} {99}},\ \bibinfo
  {pages} {233108} (\bibinfo {year} {2011})}\BibitemShut {NoStop}%
\bibitem [{\citenamefont {Otsuka}\ \emph {et~al.}(2012)\citenamefont {Otsuka},
  \citenamefont {Sugihara}, \citenamefont {Yoneda}, \citenamefont {Katsumoto},\
  and\ \citenamefont {Tarucha}}]{Otsuka:2012p081308}%
  \BibitemOpen
  \bibfield  {author} {\bibinfo {author} {\bibfnamefont {T.}~\bibnamefont
  {Otsuka}}, \bibinfo {author} {\bibfnamefont {Y.}~\bibnamefont {Sugihara}},
  \bibinfo {author} {\bibfnamefont {J.}~\bibnamefont {Yoneda}}, \bibinfo
  {author} {\bibfnamefont {S.}~\bibnamefont {Katsumoto}}, \ and\ \bibinfo
  {author} {\bibfnamefont {S.}~\bibnamefont {Tarucha}},\ }\href@noop {}
  {\bibfield  {journal} {\bibinfo  {journal} {Physical Review B}\ }\textbf
  {\bibinfo {volume} {86}},\ \bibinfo {pages} {081308} (\bibinfo {year}
  {2012})}\BibitemShut {NoStop}%
\bibitem [{\citenamefont {Studenikin}\ \emph {et~al.}(2012)\citenamefont
  {Studenikin}, \citenamefont {Aers}, \citenamefont {Granger}, \citenamefont
  {Gaudreau}, \citenamefont {Kam}, \citenamefont {Zawadzki}, \citenamefont
  {Wasilewski},\ and\ \citenamefont {Sachrajda}}]{Studenikin:2012p226802}%
  \BibitemOpen
  \bibfield  {author} {\bibinfo {author} {\bibfnamefont {S.}~\bibnamefont
  {Studenikin}}, \bibinfo {author} {\bibfnamefont {G.}~\bibnamefont {Aers}},
  \bibinfo {author} {\bibfnamefont {G.}~\bibnamefont {Granger}}, \bibinfo
  {author} {\bibfnamefont {L.}~\bibnamefont {Gaudreau}}, \bibinfo {author}
  {\bibfnamefont {A.}~\bibnamefont {Kam}}, \bibinfo {author} {\bibfnamefont
  {P.}~\bibnamefont {Zawadzki}}, \bibinfo {author} {\bibfnamefont
  {Z.}~\bibnamefont {Wasilewski}}, \ and\ \bibinfo {author} {\bibfnamefont
  {A.}~\bibnamefont {Sachrajda}},\ }\href@noop {} {\bibfield  {journal}
  {\bibinfo  {journal} {Physical review letters}\ }\textbf {\bibinfo {volume}
  {108}},\ \bibinfo {pages} {226802} (\bibinfo {year} {2012})}\BibitemShut
  {NoStop}%
\bibitem [{\citenamefont {Dial}\ \emph {et~al.}(2013)\citenamefont {Dial},
  \citenamefont {Shulman}, \citenamefont {Harvey}, \citenamefont {Bluhm},
  \citenamefont {Umansky},\ and\ \citenamefont {Yacoby}}]{Dial:2013p146804}%
  \BibitemOpen
  \bibfield  {author} {\bibinfo {author} {\bibfnamefont {O.~E.}\ \bibnamefont
  {Dial}}, \bibinfo {author} {\bibfnamefont {M.~D.}\ \bibnamefont {Shulman}},
  \bibinfo {author} {\bibfnamefont {S.~P.}\ \bibnamefont {Harvey}}, \bibinfo
  {author} {\bibfnamefont {H.}~\bibnamefont {Bluhm}}, \bibinfo {author}
  {\bibfnamefont {V.}~\bibnamefont {Umansky}}, \ and\ \bibinfo {author}
  {\bibfnamefont {A.}~\bibnamefont {Yacoby}},\ }\href@noop {} {\bibfield
  {journal} {\bibinfo  {journal} {Phys. Rev. Lett.}\ }\textbf {\bibinfo
  {volume} {110}},\ \bibinfo {pages} {146804} (\bibinfo {year}
  {2013})}\BibitemShut {NoStop}%
\bibitem [{\citenamefont {Zwanenburg}\ \emph {et~al.}(2013)\citenamefont
  {Zwanenburg}, \citenamefont {Dzurak}, \citenamefont {Morello}, \citenamefont
  {Simmons}, \citenamefont {Hollenberg}, \citenamefont {Klimeck}, \citenamefont
  {Rogge}, \citenamefont {Coppersmith},\ and\ \citenamefont
  {Eriksson}}]{Zwanenburg:2013p961}%
  \BibitemOpen
  \bibfield  {author} {\bibinfo {author} {\bibfnamefont {F.~A.}\ \bibnamefont
  {Zwanenburg}}, \bibinfo {author} {\bibfnamefont {A.~S.}\ \bibnamefont
  {Dzurak}}, \bibinfo {author} {\bibfnamefont {A.}~\bibnamefont {Morello}},
  \bibinfo {author} {\bibfnamefont {M.~Y.}\ \bibnamefont {Simmons}}, \bibinfo
  {author} {\bibfnamefont {L.~C.~L.}\ \bibnamefont {Hollenberg}}, \bibinfo
  {author} {\bibfnamefont {G.}~\bibnamefont {Klimeck}}, \bibinfo {author}
  {\bibfnamefont {S.}~\bibnamefont {Rogge}}, \bibinfo {author} {\bibfnamefont
  {S.~N.}\ \bibnamefont {Coppersmith}}, \ and\ \bibinfo {author} {\bibfnamefont
  {M.~A.}\ \bibnamefont {Eriksson}},\ }\href@noop {} {\bibfield  {journal}
  {\bibinfo  {journal} {Rev. Mod. Phys.}\ }\textbf {\bibinfo {volume} {85}},\
  \bibinfo {pages} {961} (\bibinfo {year} {2013})}\BibitemShut {NoStop}%
\bibitem [{\citenamefont {Pioro-Ladri\`{e}re}\ \emph
  {et~al.}(2007)\citenamefont {Pioro-Ladri\`{e}re}, \citenamefont {Tokura},
  \citenamefont {Obata}, \citenamefont {Kubo},\ and\ \citenamefont
  {S.Tarucha}}]{PioroLadriere:2007p024105}%
  \BibitemOpen
  \bibfield  {author} {\bibinfo {author} {\bibfnamefont {M.}~\bibnamefont
  {Pioro-Ladri\`{e}re}}, \bibinfo {author} {\bibfnamefont {Y.}~\bibnamefont
  {Tokura}}, \bibinfo {author} {\bibfnamefont {T.}~\bibnamefont {Obata}},
  \bibinfo {author} {\bibfnamefont {T.}~\bibnamefont {Kubo}}, \ and\ \bibinfo
  {author} {\bibnamefont {S.Tarucha}},\ }\href@noop {} {\bibfield  {journal}
  {\bibinfo  {journal} {Appl Phys Lett}\ }\textbf {\bibinfo {volume} {90}}
  (\bibinfo {year} {2007})}\BibitemShut {NoStop}%
\bibitem [{\citenamefont {Pioro-Ladri\`{e}re}\ \emph
  {et~al.}(2008)\citenamefont {Pioro-Ladri\`{e}re}, \citenamefont {Obata},
  \citenamefont {Tokura}, \citenamefont {Shin}, \citenamefont {Kubo},
  \citenamefont {Yoshida}, \citenamefont {Taniyama},\ and\ \citenamefont
  {Tarucha}}]{PioroLadriere:2008p776}%
  \BibitemOpen
  \bibfield  {author} {\bibinfo {author} {\bibfnamefont {M.}~\bibnamefont
  {Pioro-Ladri\`{e}re}}, \bibinfo {author} {\bibfnamefont {T.}~\bibnamefont
  {Obata}}, \bibinfo {author} {\bibfnamefont {Y.}~\bibnamefont {Tokura}},
  \bibinfo {author} {\bibfnamefont {Y.-S.}\ \bibnamefont {Shin}}, \bibinfo
  {author} {\bibfnamefont {T.}~\bibnamefont {Kubo}}, \bibinfo {author}
  {\bibfnamefont {K.}~\bibnamefont {Yoshida}}, \bibinfo {author} {\bibfnamefont
  {T.}~\bibnamefont {Taniyama}}, \ and\ \bibinfo {author} {\bibfnamefont
  {S.}~\bibnamefont {Tarucha}},\ }\href {\doibase doi:10.1038/nphys1053}
  {\bibfield  {journal} {\bibinfo  {journal} {Nat. Phys.}\ }\textbf {\bibinfo
  {volume} {4}},\ \bibinfo {pages} {776} (\bibinfo {year} {2008})}\BibitemShut
  {NoStop}%
\bibitem [{\citenamefont {Petta}\ \emph {et~al.}(2010)\citenamefont {Petta},
  \citenamefont {Lu},\ and\ \citenamefont {Gossard}}]{Petta:2010p669}%
  \BibitemOpen
  \bibfield  {author} {\bibinfo {author} {\bibfnamefont {J.~R.}\ \bibnamefont
  {Petta}}, \bibinfo {author} {\bibfnamefont {H.}~\bibnamefont {Lu}}, \ and\
  \bibinfo {author} {\bibfnamefont {A.~C.}\ \bibnamefont {Gossard}},\
  }\href@noop {} {\bibfield  {journal} {\bibinfo  {journal} {Science}\ }\textbf
  {\bibinfo {volume} {327}},\ \bibinfo {pages} {669} (\bibinfo {year}
  {2010})}\BibitemShut {NoStop}%
\bibitem [{\citenamefont {Prance}\ \emph {et~al.}(2012)\citenamefont {Prance},
  \citenamefont {Shi}, \citenamefont {Simmons}, \citenamefont {Savage},
  \citenamefont {Lagally}, \citenamefont {Schreiber}, \citenamefont
  {Vandersypen}, \citenamefont {Friesen}, \citenamefont {Joynt}, \citenamefont
  {Coppersmith},\ and\ \citenamefont {Eriksson}}]{Prance:2012p046808}%
  \BibitemOpen
  \bibfield  {author} {\bibinfo {author} {\bibfnamefont {J.~R.}\ \bibnamefont
  {Prance}}, \bibinfo {author} {\bibfnamefont {Z.}~\bibnamefont {Shi}},
  \bibinfo {author} {\bibfnamefont {C.~B.}\ \bibnamefont {Simmons}}, \bibinfo
  {author} {\bibfnamefont {D.~E.}\ \bibnamefont {Savage}}, \bibinfo {author}
  {\bibfnamefont {M.~G.}\ \bibnamefont {Lagally}}, \bibinfo {author}
  {\bibfnamefont {L.~R.}\ \bibnamefont {Schreiber}}, \bibinfo {author}
  {\bibfnamefont {L.~M.~K.}\ \bibnamefont {Vandersypen}}, \bibinfo {author}
  {\bibfnamefont {M.}~\bibnamefont {Friesen}}, \bibinfo {author} {\bibfnamefont
  {R.}~\bibnamefont {Joynt}}, \bibinfo {author} {\bibfnamefont {S.~N.}\
  \bibnamefont {Coppersmith}}, \ and\ \bibinfo {author} {\bibfnamefont {M.~A.}\
  \bibnamefont {Eriksson}},\ }\href@noop {} {\bibfield  {journal} {\bibinfo
  {journal} {Phys Rev Lett}\ }\textbf {\bibinfo {volume} {108}},\ \bibinfo
  {pages} {046808} (\bibinfo {year} {2012})}\BibitemShut {NoStop}%
\bibitem [{\citenamefont {Shi}\ \emph {et~al.}(2012)\citenamefont {Shi},
  \citenamefont {Simmons}, \citenamefont {Prance}, \citenamefont {Gamble},
  \citenamefont {Koh}, \citenamefont {Shim}, \citenamefont {Hu}, \citenamefont
  {Savage}, \citenamefont {Lagally}, \citenamefont {Eriksson}, \citenamefont
  {Friesen},\ and\ \citenamefont {Coppersmith}}]{Shi:2012p140503}%
  \BibitemOpen
  \bibfield  {author} {\bibinfo {author} {\bibfnamefont {Z.}~\bibnamefont
  {Shi}}, \bibinfo {author} {\bibfnamefont {C.~B.}\ \bibnamefont {Simmons}},
  \bibinfo {author} {\bibfnamefont {J.~R.}\ \bibnamefont {Prance}}, \bibinfo
  {author} {\bibfnamefont {J.~K.}\ \bibnamefont {Gamble}}, \bibinfo {author}
  {\bibfnamefont {T.~S.}\ \bibnamefont {Koh}}, \bibinfo {author} {\bibfnamefont
  {Y.-P.}\ \bibnamefont {Shim}}, \bibinfo {author} {\bibfnamefont
  {X.}~\bibnamefont {Hu}}, \bibinfo {author} {\bibfnamefont {D.~E.}\
  \bibnamefont {Savage}}, \bibinfo {author} {\bibfnamefont {M.~G.}\
  \bibnamefont {Lagally}}, \bibinfo {author} {\bibfnamefont {M.~A.}\
  \bibnamefont {Eriksson}}, \bibinfo {author} {\bibfnamefont {M.}~\bibnamefont
  {Friesen}}, \ and\ \bibinfo {author} {\bibfnamefont {S.~N.}\ \bibnamefont
  {Coppersmith}},\ }\href@noop {} {\bibfield  {journal} {\bibinfo  {journal}
  {Phys. Rev. Lett.}\ }\textbf {\bibinfo {volume} {108}},\ \bibinfo {pages}
  {140503} (\bibinfo {year} {2012})}\BibitemShut {NoStop}%
\bibitem [{\citenamefont {Shevchenko}\ \emph {et~al.}(2010)\citenamefont
  {Shevchenko}, \citenamefont {Ashhab},\ and\ \citenamefont
  {Nori}}]{Shevchenko:2010p1}%
  \BibitemOpen
  \bibfield  {author} {\bibinfo {author} {\bibfnamefont {S.~N.}\ \bibnamefont
  {Shevchenko}}, \bibinfo {author} {\bibfnamefont {S.}~\bibnamefont {Ashhab}},
  \ and\ \bibinfo {author} {\bibfnamefont {F.}~\bibnamefont {Nori}},\
  }\href@noop {} {\bibfield  {journal} {\bibinfo  {journal} {Physics Reports}\
  }\textbf {\bibinfo {volume} {492}},\ \bibinfo {pages} {1} (\bibinfo {year}
  {2010})}\BibitemShut {NoStop}%
\bibitem [{\citenamefont {Ribeiro}\ \emph {et~al.}(2013)\citenamefont
  {Ribeiro}, \citenamefont {Petta},\ and\ \citenamefont
  {Burkard}}]{Ribeiro:2013p235318}%
  \BibitemOpen
  \bibfield  {author} {\bibinfo {author} {\bibfnamefont {H.}~\bibnamefont
  {Ribeiro}}, \bibinfo {author} {\bibfnamefont {J.~R.}\ \bibnamefont {Petta}},
  \ and\ \bibinfo {author} {\bibfnamefont {G.}~\bibnamefont {Burkard}},\
  }\href@noop {} {\bibfield  {journal} {\bibinfo  {journal} {Phys. Rev. B}\
  }\textbf {\bibinfo {volume} {87}},\ \bibinfo {pages} {235318} (\bibinfo
  {year} {2013})}\BibitemShut {NoStop}%
\bibitem [{\citenamefont {Cao}\ \emph {et~al.}(2013)\citenamefont {Cao},
  \citenamefont {Li}, \citenamefont {Tu}, \citenamefont {Wang}, \citenamefont
  {Zhou}, \citenamefont {Xiao}, \citenamefont {Guo}, \citenamefont {Jiang},\
  and\ \citenamefont {Guo}}]{Cao:2013p1401}%
  \BibitemOpen
  \bibfield  {author} {\bibinfo {author} {\bibfnamefont {G.}~\bibnamefont
  {Cao}}, \bibinfo {author} {\bibfnamefont {H.-O.}\ \bibnamefont {Li}},
  \bibinfo {author} {\bibfnamefont {T.}~\bibnamefont {Tu}}, \bibinfo {author}
  {\bibfnamefont {L.}~\bibnamefont {Wang}}, \bibinfo {author} {\bibfnamefont
  {C.}~\bibnamefont {Zhou}}, \bibinfo {author} {\bibfnamefont {M.}~\bibnamefont
  {Xiao}}, \bibinfo {author} {\bibfnamefont {G.-C.}\ \bibnamefont {Guo}},
  \bibinfo {author} {\bibfnamefont {H.-W.}\ \bibnamefont {Jiang}}, \ and\
  \bibinfo {author} {\bibfnamefont {G.-P.}\ \bibnamefont {Guo}},\ }\href@noop
  {} {\bibfield  {journal} {\bibinfo  {journal} {Nature Comm.}\ }\textbf
  {\bibinfo {volume} {4}},\ \bibinfo {pages} {1401} (\bibinfo {year}
  {2013})}\BibitemShut {NoStop}%
\bibitem [{\citenamefont {Granger}\ \emph {et~al.}(2014)\citenamefont
  {Granger}, \citenamefont {Aers}, \citenamefont {Studenikin}, \citenamefont
  {Kam}, \citenamefont {Zawadzki}, \citenamefont {Wasilewski},\ and\
  \citenamefont {Sachrajda}}]{Granger2014preprint}%
  \BibitemOpen
  \bibfield  {author} {\bibinfo {author} {\bibfnamefont {G.}~\bibnamefont
  {Granger}}, \bibinfo {author} {\bibfnamefont {G.}~\bibnamefont {Aers}},
  \bibinfo {author} {\bibfnamefont {S.}~\bibnamefont {Studenikin}}, \bibinfo
  {author} {\bibfnamefont {A.}~\bibnamefont {Kam}}, \bibinfo {author}
  {\bibfnamefont {P.}~\bibnamefont {Zawadzki}}, \bibinfo {author}
  {\bibfnamefont {Z.}~\bibnamefont {Wasilewski}}, \ and\ \bibinfo {author}
  {\bibfnamefont {A.}~\bibnamefont {Sachrajda}},\ }\href@noop {} {\bibfield
  {journal} {\bibinfo  {journal} {arXiv preprint arXiv:1404.3636}\ } (\bibinfo
  {year} {2014})}\BibitemShut {NoStop}%
\bibitem [{\citenamefont {Assali}\ \emph {et~al.}(2011)\citenamefont {Assali},
  \citenamefont {Petrilli}, \citenamefont {Capaz}, \citenamefont {Koiller},
  \citenamefont {Hu},\ and\ \citenamefont {Das~Sarma}}]{Assali:2011p165301}%
  \BibitemOpen
  \bibfield  {author} {\bibinfo {author} {\bibfnamefont {L.~V.~C.}\
  \bibnamefont {Assali}}, \bibinfo {author} {\bibfnamefont {H.~M.}\
  \bibnamefont {Petrilli}}, \bibinfo {author} {\bibfnamefont {R.~B.}\
  \bibnamefont {Capaz}}, \bibinfo {author} {\bibfnamefont {B.}~\bibnamefont
  {Koiller}}, \bibinfo {author} {\bibfnamefont {X.}~\bibnamefont {Hu}}, \ and\
  \bibinfo {author} {\bibfnamefont {S.}~\bibnamefont {Das~Sarma}},\ }\href
  {\doibase 10.1103/PhysRevB.83.165301} {\bibfield  {journal} {\bibinfo
  {journal} {Phys. Rev. B}\ }\textbf {\bibinfo {volume} {83}},\ \bibinfo
  {pages} {165301} (\bibinfo {year} {2011})}\BibitemShut {NoStop}%
\bibitem [{\citenamefont {Petersson}\ \emph {et~al.}(2010)\citenamefont
  {Petersson}, \citenamefont {Petta}, \citenamefont {Lu},\ and\ \citenamefont
  {Gossard}}]{Petersson:2010p246804}%
  \BibitemOpen
  \bibfield  {author} {\bibinfo {author} {\bibfnamefont {K.~D.}\ \bibnamefont
  {Petersson}}, \bibinfo {author} {\bibfnamefont {J.~R.}\ \bibnamefont
  {Petta}}, \bibinfo {author} {\bibfnamefont {H.}~\bibnamefont {Lu}}, \ and\
  \bibinfo {author} {\bibfnamefont {A.~C.}\ \bibnamefont {Gossard}},\
  }\href@noop {} {\bibfield  {journal} {\bibinfo  {journal} {Phys. Rev. Lett.}\
  }\textbf {\bibinfo {volume} {105}},\ \bibinfo {pages} {246804} (\bibinfo
  {year} {2010})}\BibitemShut {NoStop}%
\bibitem [{\citenamefont {Beaudoin}\ and\ \citenamefont
  {Coish}(2013)}]{Beaudoin:2013p085320}%
  \BibitemOpen
  \bibfield  {author} {\bibinfo {author} {\bibfnamefont {F.}~\bibnamefont
  {Beaudoin}}\ and\ \bibinfo {author} {\bibfnamefont {W.~A.}\ \bibnamefont
  {Coish}},\ }\href {\doibase 10.1103/PhysRevB.88.085320} {\bibfield  {journal}
  {\bibinfo  {journal} {Phys. Rev. B}\ }\textbf {\bibinfo {volume} {88}},\
  \bibinfo {pages} {085320} (\bibinfo {year} {2013})}\BibitemShut {NoStop}%
\bibitem [{\citenamefont {Shi}\ \emph {et~al.}(2013)\citenamefont {Shi},
  \citenamefont {Simmons}, \citenamefont {Ward}, \citenamefont {Prance},
  \citenamefont {Koh}, \citenamefont {Gamble}, \citenamefont {Wu},
  \citenamefont {Savage}, \citenamefont {Lagally}, \citenamefont {Friesen},
  \citenamefont {Coppersmith},\ and\ \citenamefont
  {Eriksson}}]{Shi:2013p075416}%
  \BibitemOpen
  \bibfield  {author} {\bibinfo {author} {\bibfnamefont {Z.}~\bibnamefont
  {Shi}}, \bibinfo {author} {\bibfnamefont {C.~B.}\ \bibnamefont {Simmons}},
  \bibinfo {author} {\bibfnamefont {D.~R.}\ \bibnamefont {Ward}}, \bibinfo
  {author} {\bibfnamefont {J.~R.}\ \bibnamefont {Prance}}, \bibinfo {author}
  {\bibfnamefont {T.~S.}\ \bibnamefont {Koh}}, \bibinfo {author} {\bibfnamefont
  {J.~K.}\ \bibnamefont {Gamble}}, \bibinfo {author} {\bibfnamefont
  {X.}~\bibnamefont {Wu}}, \bibinfo {author} {\bibfnamefont {D.~E.}\
  \bibnamefont {Savage}}, \bibinfo {author} {\bibfnamefont {M.~G.}\
  \bibnamefont {Lagally}}, \bibinfo {author} {\bibfnamefont {M.}~\bibnamefont
  {Friesen}}, \bibinfo {author} {\bibfnamefont {S.~N.}\ \bibnamefont
  {Coppersmith}}, \ and\ \bibinfo {author} {\bibfnamefont {M.~A.}\ \bibnamefont
  {Eriksson}},\ }\href@noop {} {\bibfield  {journal} {\bibinfo  {journal}
  {Phys. Rev. B}\ }\textbf {\bibinfo {volume} {88}},\ \bibinfo {pages} {075416}
  (\bibinfo {year} {2013})}\BibitemShut {NoStop}%
\bibitem [{\citenamefont {Taylor}\ \emph {et~al.}(2007)\citenamefont {Taylor},
  \citenamefont {Petta}, \citenamefont {Johnson}, \citenamefont {Yacoby},
  \citenamefont {Marcus},\ and\ \citenamefont {Lukin}}]{Taylor:2007p464}%
  \BibitemOpen
  \bibfield  {author} {\bibinfo {author} {\bibfnamefont {J.~M.}\ \bibnamefont
  {Taylor}}, \bibinfo {author} {\bibfnamefont {J.~R.}\ \bibnamefont {Petta}},
  \bibinfo {author} {\bibfnamefont {A.~C.}\ \bibnamefont {Johnson}}, \bibinfo
  {author} {\bibfnamefont {A.}~\bibnamefont {Yacoby}}, \bibinfo {author}
  {\bibfnamefont {C.~M.}\ \bibnamefont {Marcus}}, \ and\ \bibinfo {author}
  {\bibfnamefont {M.~D.}\ \bibnamefont {Lukin}},\ }\href {\doibase
  10.1103/PhysRevB.76.035315} {\bibfield  {journal} {\bibinfo  {journal} {Phys.
  Rev. B}\ }\textbf {\bibinfo {volume} {76}},\ \bibinfo {pages} {035315}
  (\bibinfo {year} {2007})}\BibitemShut {NoStop}%
\bibitem [{\citenamefont {Culcer}\ and\ \citenamefont
  {Zimmerman}(2013)}]{Culcer:2013p232108}%
  \BibitemOpen
  \bibfield  {author} {\bibinfo {author} {\bibfnamefont {D.}~\bibnamefont
  {Culcer}}\ and\ \bibinfo {author} {\bibfnamefont {N.~M.}\ \bibnamefont
  {Zimmerman}},\ }\href@noop {} {\bibfield  {journal} {\bibinfo  {journal}
  {Applied Physics Letters}\ }\textbf {\bibinfo {volume} {102}},\ \bibinfo
  {pages} {232108} (\bibinfo {year} {2013})}\BibitemShut {NoStop}%
\bibitem [{\citenamefont {Morton}\ \emph {et~al.}(2011)\citenamefont {Morton},
  \citenamefont {McCamey}, \citenamefont {Eriksson},\ and\ \citenamefont
  {Lyon}}]{Morton:2011p345}%
  \BibitemOpen
  \bibfield  {author} {\bibinfo {author} {\bibfnamefont {J.~J.}\ \bibnamefont
  {Morton}}, \bibinfo {author} {\bibfnamefont {D.~R.}\ \bibnamefont {McCamey}},
  \bibinfo {author} {\bibfnamefont {M.~A.}\ \bibnamefont {Eriksson}}, \ and\
  \bibinfo {author} {\bibfnamefont {S.~A.}\ \bibnamefont {Lyon}},\ }\href@noop
  {} {\bibfield  {journal} {\bibinfo  {journal} {Nature}\ }\textbf {\bibinfo
  {volume} {479}},\ \bibinfo {pages} {345} (\bibinfo {year}
  {2011})}\BibitemShut {NoStop}%
\bibitem [{\citenamefont {Pla}\ \emph {et~al.}(2012)\citenamefont {Pla},
  \citenamefont {Tan}, \citenamefont {Dehollain}, \citenamefont {Lim},
  \citenamefont {Morton}, \citenamefont {Jamieson}, \citenamefont {Dzurak},\
  and\ \citenamefont {Morello}}]{Pla:2012p489}%
  \BibitemOpen
  \bibfield  {author} {\bibinfo {author} {\bibfnamefont {J.~J.}\ \bibnamefont
  {Pla}}, \bibinfo {author} {\bibfnamefont {K.~Y.}\ \bibnamefont {Tan}},
  \bibinfo {author} {\bibfnamefont {J.~P.}\ \bibnamefont {Dehollain}}, \bibinfo
  {author} {\bibfnamefont {W.~H.}\ \bibnamefont {Lim}}, \bibinfo {author}
  {\bibfnamefont {J.~J.}\ \bibnamefont {Morton}}, \bibinfo {author}
  {\bibfnamefont {D.~N.}\ \bibnamefont {Jamieson}}, \bibinfo {author}
  {\bibfnamefont {A.~S.}\ \bibnamefont {Dzurak}}, \ and\ \bibinfo {author}
  {\bibfnamefont {A.}~\bibnamefont {Morello}},\ }\href@noop {} {\bibfield
  {journal} {\bibinfo  {journal} {Nature}\ }\textbf {\bibinfo {volume} {489}},\
  \bibinfo {pages} {541} (\bibinfo {year} {2012})}\BibitemShut {NoStop}%
\bibitem [{\citenamefont {B{\"u}ch}\ \emph {et~al.}(2013)\citenamefont
  {B{\"u}ch}, \citenamefont {Mahapatra}, \citenamefont {Rahman}, \citenamefont
  {Morello},\ and\ \citenamefont {Simmons}}]{Buch:2013p2017}%
  \BibitemOpen
  \bibfield  {author} {\bibinfo {author} {\bibfnamefont {H.}~\bibnamefont
  {B{\"u}ch}}, \bibinfo {author} {\bibfnamefont {S.}~\bibnamefont {Mahapatra}},
  \bibinfo {author} {\bibfnamefont {R.}~\bibnamefont {Rahman}}, \bibinfo
  {author} {\bibfnamefont {A.}~\bibnamefont {Morello}}, \ and\ \bibinfo
  {author} {\bibfnamefont {M.~Y.}\ \bibnamefont {Simmons}},\ }\href@noop {}
  {\bibfield  {journal} {\bibinfo  {journal} {Nature communications}\ }\textbf
  {\bibinfo {volume} {4}},\ \bibinfo {pages} {2017} (\bibinfo {year}
  {2013})}\BibitemShut {NoStop}%
\bibitem [{\citenamefont {Yin}\ \emph {et~al.}(2013)\citenamefont {Yin},
  \citenamefont {Rancic}, \citenamefont {de~Boo}, \citenamefont {Stavrias},
  \citenamefont {McCallum}, \citenamefont {Sellars},\ and\ \citenamefont
  {Rogge}}]{Yin:2013p91}%
  \BibitemOpen
  \bibfield  {author} {\bibinfo {author} {\bibfnamefont {C.}~\bibnamefont
  {Yin}}, \bibinfo {author} {\bibfnamefont {M.}~\bibnamefont {Rancic}},
  \bibinfo {author} {\bibfnamefont {G.~G.}\ \bibnamefont {de~Boo}}, \bibinfo
  {author} {\bibfnamefont {N.}~\bibnamefont {Stavrias}}, \bibinfo {author}
  {\bibfnamefont {J.~C.}\ \bibnamefont {McCallum}}, \bibinfo {author}
  {\bibfnamefont {M.~J.}\ \bibnamefont {Sellars}}, \ and\ \bibinfo {author}
  {\bibfnamefont {S.}~\bibnamefont {Rogge}},\ }\href@noop {} {\bibfield
  {journal} {\bibinfo  {journal} {Nature}\ }\textbf {\bibinfo {volume} {497}},\
  \bibinfo {pages} {91} (\bibinfo {year} {2013})}\BibitemShut {NoStop}%
\end{thebibliography}%

\end{document}